\documentclass[10pt, journal, compsoc]{IEEEtran}

\usepackage{mystyle}

\toggletrue{complete-version}
\togglefalse{highlight-revision}

\def\doctitle{Trustworthy Distributed Certification of\\Program Execution}

\def\docauthors{Alex Wolf, Marco Edoardo Palma, Pasquale Salza, Harald C. Gall}

\def\dockeywords{%
Program execution certification, distributed computation, blockchain
}

\StrSubstitute{\doctitle}{\\}{ }[\cleandoctitle]
\StrSubstitute{\dockeywords}{.}{}[\cleandockeywords]
\hypersetup{
	pdftitle={\cleandoctitle},
	pdfauthor={\docauthors},
	pdfkeywords={\cleandockeywords}
}

\def\replicationurl{https://github.com/Lochindaal/occpReplicationPackage}
\def\monaurl{https://github.com/MEPalma/Mona/}

\DeclareAcronym{ipfs}{
	short = IPFS,
	long = {Interplanetary File System}
}
\DeclareAcronym{api}{
	short = API,
	long = {Application Program Interface}
}

\DeclareAcronym{mi}{
	short = MI,
	long = {Mona Interpreter}
}

\DeclareAcronym{jvm}{
	short = JVM,
	long = {Java Virtual Machine}
}

\DeclareAcronym{env}{
	short = env,
	long = {environment}
}

\DeclareAcronym{lifo}{
	short = LIFO,
	long = {Last-In-First-Out}
}

\DeclareAcronym{em}{
	short = EM,
	long = {Execution Mode}
}

\DeclareAcronym{expr}{
	short = Expr,
	long = {Evaluation Expression}
}

\DeclareAcronym{hr}{
	short = H\&R,
	long = {Halt and Resume}
}

\DeclareAcronym{ast}{
	short = AST,
	long = {Abstract Syntax Tree}
}

\DeclareAcronym{seqid}{
	short = seqid,
	long = {Sequence Id}
}

\DeclareAcronym{occp}{
	short = OCCP,
	long = {On-Chain Certification Protocol}
}

\DeclareAcronym{sgx}{
	short = Intel\textsuperscript{\textregistered} SGX,
	long = {Intel\textsuperscript{\textregistered} Software Guard Extensions}
}

\DeclareAcronym{kaplan_amd_2016}{
	short = SEV,
	long = {Secure Encrypted Virtualization}
}

\DeclareAcronym{tee}{
  short = TEE,
  long = {Trusted Execution Environment}
}

\DeclareAcronym{trustZone}{
	short = TrustZone,
	long = {ARM TrustZone}
}

\DeclareAcronym{eh}{
	short = \symH,
	long = {Execution Hash}
}

\DeclareAcronym{w}{
	short = W,
	long = {Worker}
}

\DeclareAcronym{dag}{
	short = DAG,
	long = {directed acyclic graph}
}

\DeclareAcronym{zkp}{
	short = ZKP,
	long = {zero-knowledge proofs}
}

\DeclareAcronym{mpc}{
	short = MPC,
	long = {secure multi-party computation}
}

\DeclareAcronym{era}{
	short = ERA,
	long = {Equivalent Register Attack}
}

\DeclareAcronym{did}{
	short = DID,
	long = {Decentralized Identity}
}

\begin{document}

\title{\doctitle}

\author{
	Alex Wolf, Marco Edoardo Palma, Pasquale Salza, and Harald C. Gall

	\IEEEcompsocitemizethanks{ %
		\IEEEcompsocthanksitem The authors are with the University of Zurich, Zurich, Switzerland. E-mail: \href{mailto:wolf@ifi.uzh.ch}{wolf@ifi.uzh.ch}, \href{mailto:marcoepalma@ifi.uzh.ch}{marcoepalma@ifi.uzh.ch}, \href{mailto:salza@ifi.uzh.ch}{salza@ifi.uzh.ch}, \href{mailto:gall@ifi.uzh.ch}{gall@ifi.uzh.ch}.
	}
}

\IEEEtitleabstractindextext{
    \begin{abstract}
Verifying the execution of a program is complicated and often limited by the inability to validate the code's correctness. It is a crucial aspect of scientific research, where it is needed to ensure the reproducibility and validity of experimental results. Similarly, in customer software testing, it is difficult for customers to verify that their specific program version was tested or executed at all. Existing state-of-the-art solutions, such as hardware-based approaches, constraint solvers, and verifiable computation systems, do not provide definitive proof of execution, which hinders reliable testing and analysis of program results. In this paper, we propose an innovative approach that combines a prototype programming language called Mona with a certification protocol OCCP to enable the distributed and decentralized re-execution of program segments. Our protocol allows for certification of program segments in a distributed, immutable, and trustworthy system without the need for naive re-execution, resulting in significant improvements in terms of time and computational resources used. We also explore the use of blockchain technology to manage the protocol workflow following other approaches in this space. 
Our approach offers a promising solution to the challenges of program execution verification and opens up opportunities for further research and development in this area. 
Our findings demonstrate the efficiency of our approach in reducing the number of program executions by up to 20-fold, while maintaining resilience against various malicious attacks compared to existing state-of-the-art methods, thus improving the efficiency of certifying program executions.
Additionally, our approach handles up to 40\% malicious workers effectively, showcasing resilience in detecting and mitigating malicious behavior. 
In the \scera scenario, it successfully identifies divergent executions even when register values and results appear identical.
Moreover, our findings highlight improvements in time and gas efficiency for longer-running problems (scaled with a multiplier of \num{1000}) compared to baseline methods. 
Specifically, adopting an informed step size reduces execution time by up to 43-fold and gas costs by up to 12-fold compared to the baseline. 
  Similarly, the informed step size approach reduces execution time by up to 6-fold and gas costs by up to 26-fold compared to a non-informed variation using a step size of \num{1000}.
\end{abstract}


}
\maketitle

\section{Introduction}
\label{sec:introduction}
  Verifying that a program execution produced the correct output given specific inputs is a fundamental challenge in software verification~\cite{sanchez_survey_2019, walfish_verifying_2015}. 
This task involves confirming that the intended code was executed having the same input, code, and output. However, various risks arise in this confirmation process: the code could be modified by malicious actors, the inputs altered, the correct output might have been generated by a different algorithm than intended, or random errors could occur due to varying configurations. 
  Unlike verifiable computing~\cite{parno_pinocchio_2013}, which focuses on providing proofs for results, verifying complete program executions introduces distinct challenges (\eg code modifications that do not reflect in the final result) due to the difficulty of ensuring the integrity of all these aspects across the entire execution path. 
Current state-of-the-art solutions (such as Parno~\etal~\cite{parno_pinocchio_2013}, Teutsch~\etal~\cite{teutsch_scalable_2019}, and Ben-Sasson~\etal~\cite{ben-sasson_scalable_2018}) fail to guarantee the authenticity of the execution comprehensively, leaving a gap in reliable verification methods.

This problem spans multiple domains where trust in program executions is critical. 
In research, reproducibility is essential for validating findings, yet replicating complex processes is often time-consuming and resource-intensive~\cite{beam_challenges_2020, strubell_energy_2019}. 
Despite reproducibility being a core principle of scientific integrity, achieving it consistently remains challenging due to various challenges, such as the significant effort, cost, and resources required, the lack of standardized methods and tools, and the necessity for a thorough understanding of the underlying code~\cite{zwaan_making_2018, vitek_repeatability_2011}.
Similarly, in industrial settings, verifying program executions is crucial, whether for confirming that software tests were genuinely conducted or for ensuring that program outputs are trustworthy.

There are several approaches (such as Parno~\etal~\cite{parno_pinocchio_2013}, Teutsch~\etal~\cite{teutsch_scalable_2019} that aim at verifying program execution.
A traditional one involves naively re-executing the entire program and carefully scrutinizing each step to confirm that the result matches the anticipated output~\cite{walfish_verifying_2015}.
However, this method is infeasible for large and complex programs that could potentially run for days or require significant computational resources.
Additionally, it cannot differentiate between programs that produce the same output but follow different execution paths, making it unable to verify the intended execution occurred. 
Moreover, when multiple stakeholders need to verify a program's execution, repeatedly re-running the process becomes infeasible. 
For instance, during software testing in a customer project, stakeholders, \ie clients, typically lack reliable methods to confirm that tests were run on the correct version of the program, 
forcing them either to acquire significant technical expertise or to trust the developers blindly~\cite{srinivasan_software_2007}. 
Long-running processes, such as batch processing tasks, further discourage re-execution due to time and resource demands~\cite{zwaan_making_2018}.

Similarly, in the industrial setting, consider a scenario where a software company delivers a critical application to a client, claiming that it has passed all required tests. 
The client must verify that the tests were conducted on the correct code version, with accurate inputs, and valid results. 
However, there are risks that the test results could be from an outdated or modified version of the software, that incorrect inputs were used, or that results were falsified.
Both examples lack a method to authenticate the program execution in a feasible way.

Previous work has suggested various approaches to tackle such challenges. For instance, hardware-based approaches, such as \acp{tee}, offer tamper-resistant processing environments running on a separated kernel, that ensure the authenticity of executed code, the integrity of runtime states and memory~\cite{costan_intel_2016, pinto_demystifying_2019, kaplan_amd_2016, sabt_tee_2015}.
\acp{tee} provide a solid foundation for secure execution and can be effectively combined with other approaches.
However, such systems require the availability of specialized hardware, which can be a significant barrier to their widespread adoption~\cite{fei_security_2021}.
Even with hardware attestation mechanisms, there is no guarantee that malicious actors would not alter results post-execution~\cite{fei_security_2021, morbitzer_severity_2021}.
Additionally, \acp{tee} are vulnerable to side-channel and cross-layer attacks, further complicating their reliability~\cite{jauernig_trusted_2020}. 
In practice, using \acp{tee} requires fully trusting a single, centralized instance to execute the code correctly, 
demanding complete confidence in that entity. 
In the previously mentioned industry scenario, while \acp{tee} could provide a controlled environment for running tests, 
they fall short of fully addressing verification concerns unless the environment is entirely trusted and safeguarded against both malicious tampering and accidental errors. 

Software-based approaches, such as constraint solvers~\cite{parno_pinocchio_2013}, do not require specialized hardware and have the theoretical capability to analyze every line of code in a given constraint.
Nonetheless, the verification process would require either a profound understanding of the internals of the code or blind trust in a third party that provides the constraints.
Additionally, when dealing with recursive functions or large programs, the number of constraints required would increase exponentially, leading to significant computational and practical challenges~\cite{vogel_challenges_2019, sanchez_survey_2019}.
In the industry scenario, verification through constraints would require detailed constraints for each code statement to ensure that the correct code, inputs, and intermediate results were processed accurately.
Therefore, the feasibility and ease-of-use of such systems remain questionable at best.

Other software-based approaches, such as verifiable computation systems, face similar challenges as constraint solvers.
Verifiers would require deep knowledge of the code to generate and verify the necessary assertions to produce a proof.
Additionally, writing assertions depends on an understanding of the code and domain, which requires significant prior knowledge, time, and effort~\cite{braun_verifying_2013}.
Alternatively, one could generate random assertions based on the execution of sample inputs, but those might not be non-trivial assertions.

A potential solution would be to combine the naive approach of re-executing the full program with a trustworthy and immutable environment to persist the result~\cite{teutsch_scalable_2019}.
By doing so, the need for multiple re-executions by every interested party would be eliminated.
However, re-executing the full program significantly reduces the usability of such a system due to increased time and computational demands, as a single re-execution could be malicious.
As such, multiple re-executions are necessary for the system to achieve trustworthiness.

By considering all the above mentioned limitations, in this paper, we present an innovative approach that combines a prototype programming language, \mona, with a certification protocol.
Our prototype language facilitates the segmentation of programs into smaller, more manageable components through our novel \ac{hr} approach. 
When combined with our certification protocol \acf{occp}, which allows these segments to be certified by re-executing them in a distributed and trustworthy manner—without resorting to naive re-execution—the protocol takes advantage of the immutability and decentralized nature of the underlying blockchain to ensure the reliable certification of program segments.
This design enhances robustness by mitigating several potential attack vectors and fortifying the system against malicious activity. 
Moreover, it can reduce the number of re-executed expressions compared to a naive re-execution of the entire program and can detect executions of different programs that produce identical register values (\ac{era}).
Our approach focuses on functional aspects, such as the authenticity of the execution given the code, memory states, and the final result of the computation, leaving non-functional properties like performance or resource consumption out of scope.
In line with other approaches in this space, we packaged our protocol utilizing a blockchain setup~\cite{teutsch_scalable_2019}, specifically \textsc{Polygon} as a layer 2 blockchain, to manage the protocol workflow. Hence, this results in a distributed, immutable, and trustworthy system~\cite{toyoda_novel_2017, hasan_combating_2019}.

We evaluate our approach by comparing the number of re-executed expressions (as it remains unaffected by parallelization) with a naive baseline approach on six popular benchmark problems. 
Thus, we evaluate the feasibility of our approach focusing on robustness and effectiveness.

To assess the effectiveness and robustness of our approach, we conducted experiments to answer three key research questions:
For \req{1} -- Program Segmentation, our experiments confirmed that the \ac{mi} prototype consistently records and replays program executions across various step sizes and benchmarks. 
This allows our prototype language to segment an execution into traces and replay any given trace to reproduce the original outcome. 
We observed a trade-off between trust and performance, where smaller step sizes enhanced trust but required more storage and computational resources, while larger step sizes improved performance at the expense of reduced trust.

For \req{2} -- Certification Protocol, we evaluated the effectiveness of \ac{occp} and found it capable of handling malicious scenarios with up to 40\% malicious workers, reliably certifying tasks or rejecting them as needed. 
Additionally, our results demonstrate that our approach can reduce the number of executed expressions by as much as 20 times.

For \req{3} -- Informed step size, our experiments show that an \textit{informed} step size reduces both time and gas costs. 
Specifically, it reduces time by up to 43-fold and gas costs by up to 12-fold compared to the baseline after a scaling multiplier of \num{1000}, with time savings already observed at a scaling multiplier of \num{100}. 
However, these reductions do not apply to gas costs for smaller scaling multipliers. 

When compared to a non-informed step size variation (see \req{2}), we observed up to a 10-fold improvement in time requirements for a step size of \num{1000}, and gas costs were reduced by up to 6-fold.

Overall, our findings demonstrate that the proposed approach reduces re-execution requirements, enables time and gas cost savings through the use of an informed step size, and exhibits robustness against various malicious attacks, achieving a zero error rate compared to the baseline.

\smallskip

\noindent
To summarize, the main contributions of this paper are:
\begin{itemize}
    \item a prototype programming language called \mona, which enables distributed and decentralized re-execution of program segments;
    \item a certification protocol \ac{occp} that allows for certification of program segments of sequential and deterministic programs in a distributed, immutable, and trustworthy system without the need for naive re-execution;
    \item an implementation of our protocol using \textsc{Polygon} as a layer 2 blockchain technology to manage the protocol workflow.
\end{itemize}
\smallskip

\noindent
The implementation, benchmark datasets, and results are available in the replication package~\cite{replicationpackage} and published at the address \url{\replicationurl}.
The \mona language~\cite{mona} is publicly available at the address \url{\monaurl}.
\smallskip

\noindent
The rest of the paper is structured as follows.
In \cref{sec:interpreter}, we provide a detailed description of our prototype language.
\Cref{sec:protocol} presents our proposed on-chain certification protocol.
\Cref{sec:experiments,sec:results} showcase the experiments, threats to validity, and results.
In \cref{sec:related_work}, we survey the related work.
Finally, we conclude in \cref{sec:conclusions} with a summary and future work.

\section{Mona Interpreter}
\label{sec:interpreter}

The system introduced in this paper utilizes a unique interpreter, built on top of ANTLR4\footnote{\url{https://www.antlr.org} v4.12.0}, to verify the reproducibility of specific segments of previous program executions, without the need to re-run the entire program from the beginning.
In more formal terms, when a program \symP is executed, it produces a sequence of expressions denoted as $\{e_0, e_1, ..., e_n\}$ along with their corresponding memory states $\{m_0, m_1, ..., m_n\}$.
In this system, the new interpreter is used to verify that evaluating \symP on a specific memory state $m$, which is obtained after evaluating some $t$ number of expressions, for another $p$ expressions, leads to the memory state $m_{t+p}$.
Importantly, this verification process is conducted without having to re-execute the entire program from $e_0$ to $e_{t+p}$.
This means that verifying $m_{t+p}$ from $m_t$ involves evaluating just $p$ expressions instead of $t+p$ expressions.
The novel \acf{hr} mechanism is what provides this functionality, which is also directly responsible for enhancing the efficiency of the proposed system.
This efficiency boost enables the system to certify program execution within a distributed setting by re-executing the program just once, overall, a crucial advancement that will be discussed in greater detail later in this paper, including its exploitation for the purpose of verifiability.

In this section, we introduce the \acf{mi} as a fully functional interpreter for the \mona programming language.
The section also discusses the innovative \acf{hr} mechanism integrated into the \ac{mi}, providing insights into its inner logic.
Additionally, an overview of the language features of the \mona language and its assumptions are presented.

\paragraph{Assumptions}
Our current implementation of \mona operates under the following assumptions:
\begin{itemize}
  \item the evaluated application operates sequentially, 
  \item it is deterministic and contains no external \acs{api} calls with non-determinism.
\end{itemize}

  While \mona is Turing-complete, the current implementation lacks certain features typically found in more mature languages. 
For example, we currently do not support multi-threading or constructs for iterating using the "in" syntax commonly seen in for-loops. 
Although for-loops are not directly available, any program requiring them can still be implemented using while-loops, which are fully supported.

\subsection{Mona Language}\label{subsec:monaLang}

\mona is a \emph{C-style}, dynamically typed, Turing-complete programming language interpreted by the \ac{mi}.
It offers support for the primitive types: \emph{character} (\eg, \code{\sq{}h\sq}), \emph{string} (as mutable \emph{lists} of characters \eg, \code{"HelloWorld"}), \emph{integer} (\eg, \code{17}), \emph{floating} (\eg, \code{1.24}), and \emph{boolean} (\code{True} and \code{False}).
Moreover, it offers native support for mutable \emph{list} types (\eg, \code{[1, \sq{}a\sq, []]}), and related functionalities for list extensions (\eg, \code{[1, 2] + [3]}), access (\eg, \code{[1, 2, 3][0]}), pythonic slicing (\eg, \code{[1, 2, 3][0:2]}, \code{[1, 2, 3][1:]}, \code{[1, 2, 3][:2]}), and item assignment (\eg, \code{[1, 2, 3][0] = 0}).
Any primitive or list values, as well as expressions, can be bounded to a variable identifier through variable declarations (\eg, \code{var identifier = 44;}).

The language exposes functionality for the definition of boolean and mathematical expressions respectively (\code{==}, \code{<=}, \code{<}, \code{>}, \code{>=}) and (\code{+}, \code{-}, \code{*}, \code{/}).
Boolean expressions can be nested in \emph{C-style} \emph{if -- else if -- else} blocks for the computation of conditional logic.
\mona supports function declarations with variable input arguments and return values.
Function invocations are considered expressions and recursive function calls are supported.
Finally, \mona offers support for loops in the form of \emph{C-style} \emph{while} expressions (\eg, \code{while (func() > min) \{print("HelloWorld!")\}}).

\subsection{Halt and Resume}

\begin{customlst}[language=Mona, float=tb, caption={An example of code in \mona language.}, label={lst:mona_example}]
decl strlst(lst) {
    if (lenof(lst) > 0) {
        print(lst[0]);
        strlst(lst[1:]);
    }
}
strlst([1, 2, 3]);
\end{customlst}

\begin{figure}[tb]
    \centering
    \includegraphics[width=1.0\linewidth]{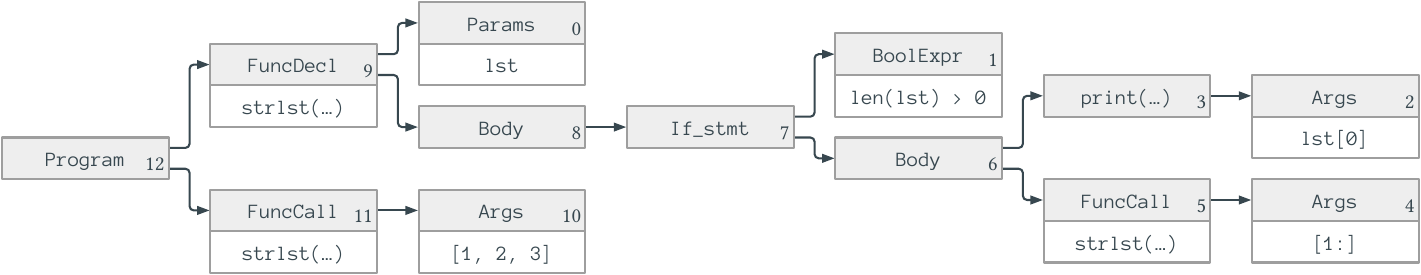}
    \caption{The \ac{ast} of the code in \cref{lst:mona_example}.}
    \label{fig:ast}
\end{figure}

Support for the \acf{hr} strategy is integrated throughout various phases of a \mona program's lifecycle.
This lifecycle encompasses several stages, beginning with the initial parsing phase, referred to as \emph{Parse-time} followed by the execution of the program, during which the evolution of the memory state is recorded in \emph{Record-time} and concluding with partial replay during \emph{Replay-time}.
During \emph{Parse-time} the \ac{mi}'s parser undertakes custom logic to construct the program's \ac{ast} while incorporating essential annotations to facilitate \ac{hr} support.
\emph{Record-time} signifies the phase in which the interpreter evaluates the program.
Here, after processing a user-defined number of program expressions, execution is halted, and a representation of the memory state at that point is persisted.
In the final phase of the program's lifecycle, known as \emph{Replay-time} the \ac{mi} is capable of loading a specific memory state recorded during \emph{Record-time}.
It then resumes program evaluation for a specified number of expressions, subsequently halting and presenting the resulting memory state.

This section presents the \acf{hr} strategy and how it seamlessly integrates into the \emph{Parse-time}, \emph{Record-time}, and \emph{Replay-time} stages of a \mona program's lifecycle.

\paragraph{Parsing Mona Programs}
During \emph{Parse-time}, the \ac{mi}'s parser employs specialized algorithms to construct the program's \ac{ast} while simultaneously adding vital annotations essential for supporting the \ac{hr} strategy.
This annotation process establishes a precise \emph{one-to-one} relationship between program expressions and a unique set of positive integers, known as \acp{seqid}.
Each program expression is assigned a specific \ac{seqid}, and the assignment is carried out in such a way that sorting the expressions based on their corresponding \ac{seqid} values naturally aligns with the order in which these expressions are evaluated.
The \ac{mi} achieves this binding through a bottom-up parsing approach, systematically assigning \ac{seqid} values to expressions from an ascending integer counter.

The \ac{seqid} annotation seamlessly integrates into the \ac{ast}, irrespective of parsing strategy (bottom-up or top-down). While the choice of bottom-up parsing was for implementation convenience, it does not limit the applicability of the \ac{seqid} annotation.

\Cref{fig:ast} presents a simplified representation of the original \ac{ast} derived from the \mona program in \cref{lst:mona_example}, showcasing the outcomes of preprocessing.
This program encompasses a function declaration and a function call designed to print the content of a string literal to the standard output.
Notably, each expression within the program is annotated with a unique \ac{seqid} value, as displayed in the bottom right corner of each grammar rule node in \cref{fig:ast}.

The significance of these \ac{seqid} values becomes apparent when we consider the execution logic within the code block labeled \code{strlst}.
The evaluation of the function in question hinges on the systematic evaluation of its various subcomponents, strictly adhering to the order dictated by their respective \ac{seqid} values.
For instance, in the function \code{strlst} (labeled as \code{9}), the evaluation initiates with the function's parameters (\code{Params}, \ac{seqid} \code{0}).
Subsequently, it proceeds to evaluate the function's body (\code{Body}, \ac{seqid} \code{8}).
However, this body is evaluated only after its child \emph{if} expression (\code{If\_stmt}, \ac{seqid} \code{7}) has been processed.
In turn, the \code{If\_stmt} itself is evaluated following the evaluation of its Boolean expression (\code{BoolExpr}, \ac{seqid} \code{1}), and subsequently, its own body (\code{Body}, \ac{seqid} \code{6}), before ultimately returning a result.
Expanding this logical sequence to the two expressions within the body leads to the following order of evaluations, where \enquote{evaluated} means that the evaluation has either returned or terminated:
\code{Params} (\ac{seqid} \code{0}),
\code{BoolExpr} (\ac{seqid} \code{1}),
\code{Args} (\ac{seqid} \code{2}),
\code{print} (\ac{seqid} \code{3}),
\code{Args} (\ac{seqid} \code{4}),
\code{FuncCall} (\ac{seqid} \code{5}),
\code{Body} (\ac{seqid} \code{6}),
\code{If\_stmt} (\ac{seqid} \code{7}),
\code{Body} (\ac{seqid} \code{8}),
\code{FuncDecl} (\ac{seqid} \code{9}).

This initial step is crucial for enabling the subsequent \emph{Resume} logic, which intelligently prunes the execution of previously evaluated expressions when loading a specific memory state, contributing to the efficiency of the \ac{hr} strategy.

\paragraph{Recording Mona Programs}
During the program execution, the \ac{mi} introduces additional operations compared to conventional interpreters.
These operations are designed to modify the memory state in a manner that allows future \emph{Replay} workflows to resume evaluations from the precise expression where the execution was halted, all while bypassing the re-execution of previously processed expressions.
The primary objective of this stage in the program's lifecycle is to ensure that the resumption logic operates seamlessly without awareness that a portion of the evaluation tree is being pruned as already executed.
To achieve this, the \emph{Record} stage introduces a novel and essential component to the environment's value access and update logic, known as the \emph{memvar}.

The memory state utilized by the \ac{mi} can be referred to as the tuple $\left(S, M, C, i\right)$.
In this tuple $S$ can be considered as a traditional program stack, hence a \ac{lifo} data structure that serves in exchanging inputs and output between expressions in the same program scope.
$M$ is the program's memory, in which the program can read and write key-value pairs, in accordance with program scopes and access policies.
$C$ is instead the list of the last executed expression \acp{seqid} for each open program scope.
Finally, $i$ is the integer index value determining which position in $C$ the program is currently evaluating.

The information contained in $C$ is crucial for \emph{Resume} workflows, as it informs the \ac{mi} about to what depth the evaluation tree was evaluated for each program scope.
In more practical terms, during evaluation, after each program expression has been evaluated and returns, its \ac{seqid} is set as the \ac{seqid} in $C$ for the current scope index $i$.
For instance, let us consider the program illustrated in \cref{lst:mona_example}.
At the outset of evaluation, the values of $C$ and $i$ are set to $\left[-1\right]$ and $0$ respectively, indicating that the program's main scope has not yet executed any expressions.
As the program begins, it encounters the first function call, \code{strlst([1, 2, 3])}.
Initially, it evaluates the argument expression, \code{Args([1, 2, 3])}.
Upon completion, this update causes $C$ to become $[10]$.
Furthermore, this function call introduces a new scope, where \code{strlst} is evaluated.
Consequently, the values of $C$ and $i$ change to $\left[10, -1\right]$ and $1$ respectively, signifying that the program is now operating within scope $1$, where no expressions have been executed.
The \code{If\_stmt} in the function's body causes another scope addition whenever the evaluation unfolds in its body.
As the boolean expression \code{BoolExpr} evaluates to \code{True} in the current scope, $C$ and $i$ are updated by \code{If\_stmt} to $\left[10, 1, -1\right]$ and $2$ before evaluating its body.
Continuing with the example, as the evaluation of \code{strlst} proceeds and computes the first print expression, $C$ and $i$ become $\left[10, 1, 3\right]$ and $2$ respectively.
As the recursive call to \code{strlst} is executed, it updates $C$ and $i$ to $\left[10, 1, 4, -1\right]$ and $3$, signaling another scope addition.
When scopes are eventually closed, as in the case of a return from a call to \code{strlst}, the last open scope is removed from $C$, and $i$ is decremented by one step. Consequently, the program concludes with $C$ and $i$ holding the values $[12]$ and $0$ respectively, reflecting the scope changes and expression execution throughout the program's evaluation.

While using $C$ and $i$ to keep track of the evaluated tree depth helps the \emph{Resume} logic determine when to prune evaluation, it is insufficient for handling scenarios where the unfolding of evaluation depends on execution data.
For instance, \emph{If Blocks} contain multiple expression bodies, of which only one should be executed based on the boolean expression of each \emph{if/else if} conditions.
Similar limitations are found in other \emph{C-Style} constructs like loops, where the body is executed based on loop conditions and update expressions.
To address such cases, we introduce the concept of \emph{memvars}.
This novel strategy connects the program's evaluation components with the program's environment and caches associations between evaluation components and stack values.
This ensures that the evaluation always receives the expected value during both execution and resumption.
During execution, when a component needs to retrieve the value of a branch decision variable, it invokes the \emph{memvar} to read the stack.
To do it, the component provides its \ac{seqid} and a unique identifier representing the name of the variable being read.
This identifier is most relevant when a component accesses multiple such values in its logic thereby ensuring multiple \emph{memvar} do not overwrite each other.
The \emph{memvar} uses these values to search in $M$ within the current evaluation scope for the cached output of the variable.
If available, it is simply returned to the component; otherwise, the value is fetched from $S$, stored in $M$, and then returned.
As a result, during evaluation, \emph{memvar} continually caches values from $S$ into $M$.
However, during \emph{Resume}, these values are retrieved from the cache rather than from $S$ when pruning.
This strategy ensures that when resuming, the evaluation unfolds in the correct branch of the sub-program without requiring explicit support within the evaluation components.
Importantly, this strategy does not lead to memory leaks in $M$.
When a scope is fully evaluated and closed, it is not only reflected in $C$ but also in $M$, where all data-related values from the just-closed scope are removed.
The \emph{memvar} only retains values that certain evaluation nodes may have used within the currently open evaluation scopes.

While the \ac{mi} handles the \ac{seqid} update logic and \emph{memvar} strategies during program execution, it also takes responsibility for creating periodic snapshots of the execution memory.
The \ac{mi} creates periodic snapshots of the execution memory as part of its program evaluation process.
To achieve this, it maintains a counter that keeps track of the number of expressions executed.
When this counter reaches a user-defined threshold, the evaluation is paused, and the current memory state is saved to disk.
The threshold is denoted as the \emph{Step} value, which infers the number of expressions evaluated between \emph{Halt} events.
It is worth noting that program start and end snapshots are exceptions and are always recorded, regardless of the expression count.
Moreover, unlike the relatively coarse-grained node separation seen in the \ac{ast} in \cref{fig:ast}, the \ac{mi} produces a significantly more fine-grained node separation.
This means that the program can be halted in a much larger number of scenarios, including during the evaluation of expressions.

\paragraph{Replaying Mona Programs}
In the final phase, known as \emph{Replay-time}, the \ac{mi} can be directed to load a specific memory state recorded during the \emph{Record-time} phase.
It then proceeds to \emph{resume} program evaluation from the saved memory state, continuing for a specified number of expressions.
During this \emph{Resuming} action, the \ac{mi} traverses the \ac{ast} but prunes previously evaluated subtrees, utilizing \acp{seqid} to determine which parts of the evaluation tree should be skipped.

Thanks to the work performed by the \ac{mi} during \emph{Record-time}, the \emph{Resume} logic presents no significant deviations from traditional evaluation logic.
This means that the \ac{mi} evaluates the program without being aware that it is resuming from a non-clean memory state.
Each evaluation node follows a logic wherein, if the value of $C$ at $i$ is greater than its own \emph{seqid}, it will not execute its inner logic but will simply return, effectively pruning its subtree of evaluation.
This logic is valid because if the environment's active \ac{seqid} is greater than that of a node, it indicates that this node has already been evaluated.
Simultaneously, \emph{memvar} ensures that when a critical branching value is being read, it is always available.

As a result, the \emph{Resume} logic can reconstruct the evaluation tree by loading the provided memory state into memory and resetting $i$ to $0$.
The previously executed evaluation tree is pruned, and execution resumes whenever the environment's active \ac{seqid} is not greater than the current node being evaluated.

\section{\acf{occp}}
\label{sec:protocol}

\ac{occp} combines a certification mechanism with blockchain technology to provide an immutable, decentralized, and trustworthy system for verifying program execution.
To enable distributed certification, we leverage the functionality of \acf{hr} of \ac{mi} (see \cref{sec:interpreter}).
This feature is essential for splitting program execution into parts or traces that can be independently and in parallel managed by a distributed and trustworthy system.
A centralized approach to computing the entire program is unfeasible because it cannot guarantee that every participant will execute their part correctly.
With the ability to \ac{hr} execution, the interpreter enables the re-execution of all expressions once, as in the initial and original execution of the program, but with the added advantage of parallelization, which reduces the probability of introducing malicious behavior.

The certification of program execution is essentially the problem of connecting each sub-execution of the program to its successor in a distributed manner.
We can think of this process as the workers trying to solve a puzzle by finding the correct sequence of sub-executions.
To achieve this, the workers in the blockchain provide the output hash of each sub-program.
With at least half of the workers in agreement on the output hash, the sequence of the execution memory states can be rebuilt.
If it is possible to rebuild the sequence, the sequence is hashed and compared to the one provided by the original executor.
However, if a sequence cannot be rebuilt due to weak consensus among workers regarding trace connections, the corresponding traces are re-evaluated until agreement is reached.
Ultimately, the workers either agree that a correct execution sequence has been reconstructed or agree that the given program cannot produce a sequence for the execution traces provided.

In this section, we present \ac{occp} that we have developed for the distributed certification of program execution on blockchain technology.
We describe the actors and workflow involved in the protocol, along with how it copes with possible malicious attacks.
We also present the implementation details of \ac{occp} using Polygon as layer 2 blockchain technology.

\subsection{Actors and Workflow}\label{subsec:actwork}
\begin{figure*}[tb]
    \centering
    \includegraphics[width=0.95\linewidth]{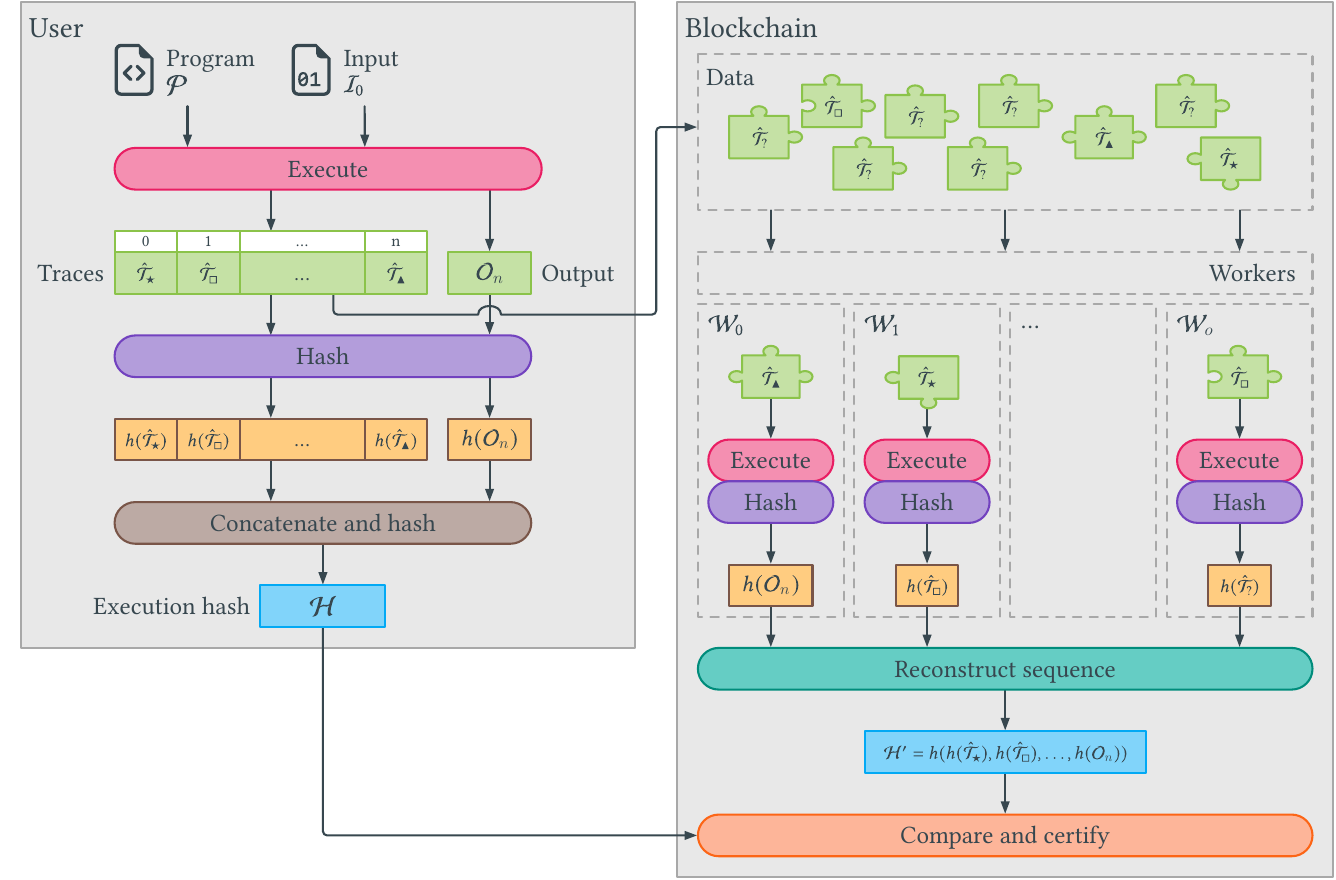}
    \caption{An overview of the interaction between the user operations and protocol on the blockchain.}
    \label{fig:overview}
\end{figure*}

A simplified overview of how the approach works is illustrated in \cref{fig:overview}.
We define a \emph{User} as the entity who wishes to verify the execution of a program.
In particular, here we refer to a program \symP as an evaluable \mona language derivation.
In order to verify an execution of \symP, the \emph{User} is expected to execute the \acf{mi} in \emph{Record Mode} on \symP and produce the set of execution \emph{Trace}s \symT.
This means that the  \ac{mi} is given a fixed and arbitrary number of execution steps after which a \emph{Halt} event is invoked; or in other words the number of expressions between two \emph{Halt} events.
Each \emph{Halt} event produces a \symT, which is the tuple $\left(\mathI_t,\ s,\ \mathP,\ \mathO_t\right)$ where $\mathI_t$ is the program's memory state at time $t$, and $\mathO_t$ is the program's memory obtained after executing $P$ from $\mathcal{I}_t$ for $s$ steps.
It is important to note that the recording of the program execution results in an ordered set of $\mathT: \{\mathT_i\}$.
It is therefore true that $\mathI_{t+1} \equiv \mathO_{t}$.
Moreover, the step value $s$ is guaranteed to be constant for all traces but the trace for which $\mathO_t$ is the program's final memory state, for which $s$ is equal to the remaining number of steps from the penultimate trace.

After the execution recording, the user provides the system with an \ac{eh}.
It encodes the sequence of memory states encountered during the execution.
This is computed by first applying the hashing function $h$ on the sequence $\left(\mathI_{0}, \mathO_{0}, \mathO_{1}, \dots, \mathO_{n}\right)$ where $n$ is the total number of traces.
The resulting hashed values sequence $\left(h_0, h_1, \dots, h_n\right)$ is applied again to \emph{h} to obtain \symH.
Hence, from $\mathT$, the \emph{User} produces $\hat{\mathT}$, for which $\hat{\mathT_i} = \mathT_i \setminus \mathO_i$ and sends the system a certification request consisting of the tuple $(\hat{\mathT}, \mathH)$, as illustrated in the \emph{User} section in \cref{fig:overview}.
Thus, the final state $\mathO_{n}$ is not directly passed to the replay phase.
The system verifies the execution by delegating to the blockchain's \acfp{w}, the task of computing $\mathO_i$ for each $\mathT_i$, and obtaining the same \symH as the one provided.
More specifically, for each trace $\hat{\mathT}_i$ a worker will compute the resulting memory state $\mathO_i$ by resuming \symP for $s$ expressions from $\mathI_i$ using the \ac{mi}.

\begin{figure*}[tb]
    \centering
    \includegraphics[width=0.95\linewidth]{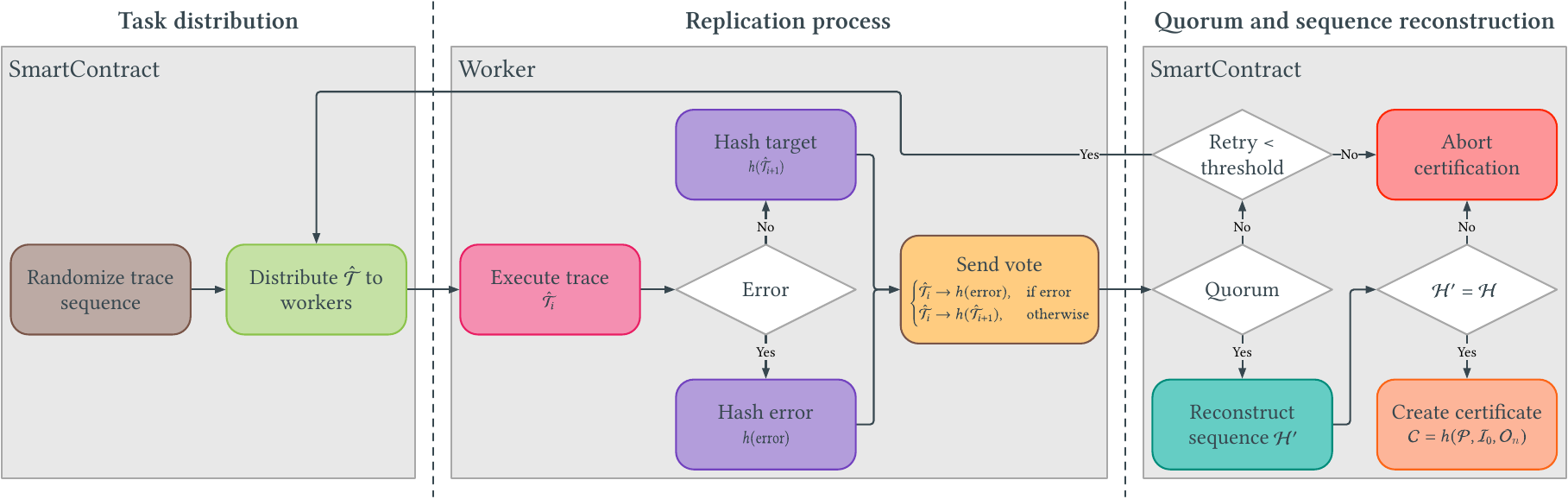}
    \caption{The detailed flowchart of \ac{occp}.}
    \label{fig:occp}
\end{figure*}

The following concepts regarding the blockchain side are detailed in the flowchart depicted in \cref{fig:occp}.
Once a worker obtains $\mathO_i$, it replies to the system with the hash of $\mathO_i$ $h_i$.
This reply represents a \emph{vote}, for which $\mathI_i$ is the source memory state for the output state descried by $h_i$.
Therefore, after the workers voted for each $\mathI_i$, the attempts to reconstruct the execution's memory state sequence.
Such computation is carried out by the workers until quorum is reached on the sequence value, \ie of all the responses, the sequence computed by half of the workers plus one, with a simple majority rule implemented to improve decision-making efficiency and responsiveness, while allowing system operators to configure the majority threshold in the smart contract based on specific needs~\cite{mossel2014majority,chen2005majority,Mukhopadhyay2020VoterAM,buchanan1961simple}.
A \ac{w} builds such a sequence by first computing to which $h_i$ the workers agree $\mathI_i$ should map to.
We call $h_{qi}$ the output hash with quorum votes for some $\mathI_i$.
Hence, each $\mathI_i$ is connected to some other $\mathI_p$ for which $h_{qi}$ equals to the hash value of $h_p$.
If a sequence going through each $\mathI_i$ can be computed, then the system certifies that for this sequence the same hash value \ac{eh} as the one submitted by the \emph{User} can be computed.
Instead, should a complete sequence not be computable, the workers provide the system with the set of traces that do not reach quorum to some output memory state.
Thus, the system delegates the computation of such traces to the workers, until a quorum is obtained.
Additionally, the system ensures that such traces are not delegated to the workers who worked on the traces previously.
Therefore, ensuring that conflicts are resolved by a diverse set of workers.
Specifically, each worker is assigned a unique identifier through which we distinguish the workers to ensure this.
Every worker is viewed equivalently by the protocol and not further distinguished based on other criteria.

Therefore, we can think of the traces in the form of a puzzle that needs to be solved by the workers.
More specifically, the puzzle consists of a very specific \ac{dag}, which is structured as a linear chain\textemdash a \emph{straight line}. Formally, the \ac{dag} is defined as $G = \left(V, E\right)$ with $N$ vertices, each representing a trace $\hat{\mathT}_i$.
The goal of the puzzle is to find the correct sequence of connecting the vertices to produce a desired final output $\mathO_n$ and generating a sequence hash \symH that matches the one provided by the \textit{user}.
Therefore, each \ac{w} proposes independently an edge by computing the resulting vertex from the one it got assigned.

Due to obfuscating the target of each trace $\hat{\mathT}_i$ and ensuring that the correct sequence of traces is unknown to the workers, the only possibility of certifying an execution is through re-execution and matching the newly computed outputs to reconstruct the sequence hash \symH as illustrated by \cref{fig:overview}.

However, before we reconstruct the sequence, we execute conflict checks.
Any vertex that produces a conflict and is therefore not a vertex that has quorum is redistributed to a worker.
To formalize this, we first define the following notions.

We have a finite set of vertices, denoted as $V = \{v_1, v_2, \ldots, v_N\}$, where each $v_i$ represents a trace.
Additionally, we have a set of directed edges, $E$, where each edge is represented as an ordered pair $(v_i, v_j)$, indicating a directed edge from trace $v_i$ to trace $v_j$.
These edges represent the voting actions of workers.
Furthermore, the edges in $E$ form a linked list, where each trace is connected to exactly one other trace in the list, with the exception of the last trace, which has no outgoing edge.
Formally, for each $v_i$, there exists at most one $v_j$ such that $\left(v_i, v_j\right) \in E$, and for the last trace $v_N$, there exists no $v_j$ such that $\left(v_N, v_j\right) \in E$.
  While the final output $\mathO_{n}$ is excluded from the set of traces $\hat{\mathT}$, the execution of the final trace by the workers would create a new edge connecting to the final output $\mathO_{n}$, resulting in $|\hat{\mathT}|$ edges. 
For instance, consider $\hat{\mathT} = {\hat{\mathT}_0, \hat{\mathT}_1, \hat{\mathT}_2}$, where the correct sequence is $\hat{\mathT}_0 \rightarrow \hat{\mathT}_2 \rightarrow \hat{\mathT}_1 \rightarrow \mathO_n$. 
Re-execution of $\hat{\mathT}_1$ would generate the edge $\left(\hat{\mathT}_{1}, \mathO_n\right)$, leading to the edge set $E = {\left(\hat{\mathT}_{0}, \hat{\mathT}_{2}\right), \left(\hat{\mathT}_{2}, \hat{\mathT}_{1}\right), \left(\hat{\mathT}_{1}, \mathO_n\right)}$. 
  Consequently, a valid voting sequence must satisfy $|\hat{\mathT}| = |E|$. In cases of conflicting votes, only edges that achieve quorum are considered valid.

We define a \emph{correct execution} as one that satisfies two conditions: 
\begin{inparaenum} 
\item it matches the expressions provided by the user, and 
\item it can be reproduced by the workers through the re-execution of traces, resulting in a linear sequence of traces.
\end{inparaenum}

\begin{lem}
    \label{lemma1}
    A correct execution results in a \ac{dag} represented as a straight line, 
    where $|\text{Traces}| = |\text{Edges}|$.
\end{lem}

This formalization clarifies the structure and relationships among the traces, thereby aiding in the analysis and verification of the votes. 
These checks act as safeguards for \cref{lemma1}, ensuring that before the \ac{occp} reconstructs the sequence, the votes on the traces are evaluated to determine if they lead to a potentially valid sequence.

\paragraph{Check 1: No vertex is allowed to have more than one outgoing connection}
For each vertex $v_i$ representing a trace, there should exist at most one $v_j$ such that $\left(v_i, v_j\right)$ represents an outgoing connection:
\begin{math}
    \forall v_i \in V: \left|\{\left(v_i, v_j\right) \in E\}\right| \leq 1
\end{math}.
This check ensures that each trace, except the last one, has at most one outgoing connection to the next trace.

\paragraph{Check 2: No vertex is allowed to have more than one incoming connection}
Similarly, for each vertex $v_i$ representing a trace (except the first vertex), there should exist at most one $v_j$ such that $\left(v_j, v_i\right)$ represents an incoming connection:
\begin{math}
    \forall v_i \in V: \left|\{\left(v_j, v_i\right) \in E\}\right| \leq 1
\end{math}.
This check ensures that each trace (except the first) has, at most, one incoming connection.

\paragraph{Check 3: There should be no cycle in the graph}
To maintain the acyclic nature of the graph, we perform cycle detection.
\begin{math}
  \forall \left(v_i, v_j\right) \in E, \left|\{v_j \rightsquigarrow v_i\}\right| < 1
\end{math}.
Where $v_j \rightsquigarrow v_i$ means there exists a directed path from $v_j$ to $v_i$. 
This condition ensures that there is no way to return to a vertex $v_i$ starting from its neighbors $v_j$, \ie, no cycles exist. 
\smallskip

Note that quorum is considered for all of the checks. Specifically, if an edge receives multiple conflicting votes from workers, the validation checks will take into account the majority vote or the agreed-upon decision by the quorum.
It means that if an edge is voted upon more than once, and the votes conflict, the other conflicting votes will be disregarded in favor of the majority vote.
This approach promotes the resilience of the protocol to discrepancies and malicious actors and ensures that decisions regarding edge validity are based on a collective consensus.
Therefore, Checks 1, 2, and 3 collectively safeguard the fulfillment of \cref{lemma1} and ensure that the resulting graph is indeed a \ac{dag}.

\paragraph{Voting and quorum example}
To further clarify the voting and quorum mechanism, consider the scenario where four workers, ${W_0, W_1, W_2, W_3}$, are tasked with reconstructing a set of traces ${\hat{\mathT}_0, \hat{\mathT}_1, \hat{\mathT}_2}$. 
For simplicity, we omit hashing details and refer directly to the traces. 
Initially, each worker is randomly assigned a trace: $W_0$ works on $\hat{\mathT}_0$ and votes that re-executing this trace leads to $\hat{\mathT}_2$; $W_1$ votes that $\hat{\mathT}_1$ results in output $O$; and $W_2$ votes that $\hat{\mathT}_2$ leads to $\hat{\mathT}_1$. 
This voting yields a trace sequence $\hat{\mathT}_0 \rightarrow \hat{\mathT}_2 \rightarrow \hat{\mathT}_1 \rightarrow O$, where each trace currently has quorum and is ready for verification.

Now, consider an alternative voting sequence where $W_0$ votes that $\hat{\mathT}_0$ leads to $\hat{\mathT}_1$, while the other workers vote as before. 
In this case, $\hat{\mathT}_1$ receives two incoming edges, violating \textit{Check 2} of \cref{lemma1}. 
Consequently, the \ac{occp} reassigns $\hat{\mathT}_0$ and $\hat{\mathT}_2$ to other workers. 
Worker $W_3$ re-executes $\hat{\mathT}_0$ and votes that it leads to $\hat{\mathT}_2$, while $W_1$ re-executes and votes that $\hat{\mathT}_2$ leads to $\hat{\mathT}_1$. 
This results in quorum with two votes for $\hat{\mathT}_2 \rightarrow \hat{\mathT}_1$. 
However, this still leaves two outgoing votes from $\hat{\mathT}_0$, violating \textit{Check 1}. 
Therefore, $\hat{\mathT}_0$ must be re-executed. 
Achieving quorum for all traces necessitates a worker voting for $\hat{\mathT}_0 \rightarrow \hat{\mathT}_2$.

\subsection{Malicious Attacks Protection}\label{subsec:prot:malProtec}

The proposed \ac{occp} protocol is designed to mitigate the impact of malicious actors who may try to undermine the system's trustworthiness.
Here are some scenarios in which the protocol can protect against malicious behavior.

\paragraph{Malicious workers intentionally return incorrect memory states}
In this scenario, workers may try to tamper with the memory states they return to the system.
The protocol can detect such behavior by comparing the sequence hash \symH to the constructed one.
If they do not match, the system can reassign the task to another worker, and it will be ignored if quorum is reached on another correct result (see also \sclaz case in \cref{sec:experiments}).

\paragraph{Malicious users submit an incorrect \ac{eh}}
In this scenario, a user may try to submit an incorrect \ac{eh} to trick the system into certifying an incorrect execution.
However, the protocol can detect this behavior because the sequence of memory states is included in the \ac{eh}.
If the sequence does not match the one computed by the workers, the system will not certify the execution (see also \scmal case in \cref{sec:experiments}).

\paragraph{\acf{era}}
In this scenario, a malicious user may try to submit a different program than the one that was executed.
This scenario showcases that programs producing the same register values and results at each step can be differentiated by the system.
Furthermore, this scenario identifies a gap in existing approaches that produce proofs based on these register values~\cite{ben-sasson_scalable_2018,ben-sasson_succinct_2014,khaburzaniya_aggregating_2021}.
Hence, it is possible to provide a different implementation that results in the same register states and outcome, rendering such systems unable to provide verification of executions with certainty and ease-of-use.

To further illustrate this case consider two research teams implementations:
\begin{inparaenum}
    \item the first one using a recursive approach,
    \item while the other one reads a pre-defined sequence from an array.
\end{inparaenum}
  While both algorithms produce the same register states and outputs they differ in many other aspects (\eg code, execution trace).
Therefore, generating proofs using register states and outputs would not guarantee that a specific algorithm was used to compute said outputs.

The protocol can detect such behavior through the combination of re-executed traces by the workers that vote on the correct path and the fact that \mona records the flow of an application through an \ac{ast} traversal mechanism.
Additionally, \mona records further values (\eg, memory, registers, IO) that make the execution unique.
Furthermore, the sequence hash \symH includes those further values and would lead to a different hash.
Thus, replacing any trace or part of the program would be noticed by the workers and the reconstructed sequence hash would be distinguishable from the originally committed one.

\paragraph{Malicious workers collude to manipulate the outcome}
In this scenario, a group of workers may collude to manipulate the outcome of the quorum voting process by intentionally returning incorrect memory states.
However, the protocol can detect such behavior by relying on the nature of blockchain networks, \ie, we rely on the fact that it is highly unlikely that all the workers are colluding.
Therefore, quorum on a sequence must be reached and additionally the sequence hash $H$ must be reproduced for a certificate to be issued.

\paragraph{Malicious workers collude with the User to manipulate the outcome}
In this scenario, similarly to the colluding workers case, the workers may collude with the User to manipulate the outcome of the quorum voting process.
Analogous to the previous case, we rely on the nature of the blockchain network to deal with this issue.

\paragraph{False Negatives and Positives}
Considering the different possible malicious attacks we discuss the impact and likelihood of false negatives and positives. 

  A false negative occurs when malicious workers produce a non-conflicting trace sequence that aligns with a hash $\mathH$ different from the intended one. 
The probability of generating such a valid non-conflicting sequence across the large number of possible graph configurations is low and hinges on achieving a malicious majority vote. 
The protocol also requires a consistent quorum across traces to finalize certification, making it difficult for adversaries to succeed.

To illustrate the probability of producing a non-conflicting sequence when a majority vote of colluding malicious workers is achieved, consider the following. 
  Let $n$ denote the number of distributed traces $|\hat{\mathT}|$.
The number of possible connected graphs using $n$ traces is $2^{{n}\choose{2}}$, while the number of valid non-conflicting sequences is $n!$. 
The probability of randomly guessing one of these non-conflicting sequences from all possible configurations is $p_{nonConflicting} = \frac{n!}{2^{{n}\choose{2}}}$. 
However, even if this sequence is found, the success of malicious actors still depends on reaching a majority vote.

Additionally, for a false negative to occur, malicious workers must consistently generate non-conflicting trace sequences across rounds until a threshold $t$ is met. 
The probability that malicious workers achieve quorum for all traces $\hat{\mathT}$ in the first round, assuming each trace is controlled by a malicious worker, is $p^{|\hat{\mathT}|}$, where $p$ represents the likelihood of a worker being malicious. 
As the number of traces increases (with smaller step sizes), this probability decreases.

In scenarios where malicious workers collaborate effectively, assuming they operate without internal conflicts and can communicate their intended trace sequences, they must control all assigned traces to avoid conflicts. 
As workers are excluded from revoting on the same trace, the group with a larger number of members is more likely to dominate the outcome.
The probability of malicious workers controlling all traces depends on the number of traces, which is influenced by the step size. 
A smaller step size increases the number of traces, reducing the chance that a majority of malicious workers are assigned all of them. 
Conversely, a larger step size decreases the number of traces, increasing the likelihood of malicious control.
For instance, if the number of distributed traces is smaller than the total population of workers, the likelihood of malicious workers receiving all traces increases.
Overall, a false negative would occur in scenarios where malicious workers consistently produce a non-conflicting and incorrect sequence hash, reaching quorum and generating a reconstructed sequence hash $\mathH'$ that deviates from the original $\mathH$ once the threshold $t$ is met.

False positives arise when malicious workers either guess a sequence resulting in a hash collision for SHA-256 or collaborate as a group with a malicious user. 
In the latter scenario, given a majority, the conditions mirror those faced by non-malicious workers cooperating with an honest user.

The complexities arising from the voting mechanism\textemdash~such as exclusions and varying communication strategies\textemdash~warrant more detailed analysis in future research.

Our approach focuses on preventing false positives, as these have a higher impact as this would entail a falsely certified execution. On the other hand, a false negative necessitates a renewed re-execution. 

\subsection{Implementation on Polygon layer 2 blockchain}\label{subsec:prot:implBC}
We have implemented a \smartcontract (\num{382} LOC) that utilizes the \ac{occp} to manage and store all traces $\hat{\mathT}$ for each program \symP uploaded.
The \smartcontract was implemented by using the \emph{Solidity} language\footnote{\url{https://soliditylang.org/}: v0.8.2}.
However, in order to improve the efficiency and scalability of our protocol, we propose an adaptation to a layer 2 \textsc{Polygon} chain.
By leveraging the benefits of \textsc{Polygon}, we can significantly reduce gas fees and increase the transaction throughput for our \smartcontract.
Additionally, leveraging \textsc{Polygon} as a layer 2 chain, we can store the final certificate permanently on the underlying \textsc{Ethereum} layer 1 chain once the main operations of the protocol have been completed.
This approach ensures that the certificate is permanently stored on a more secure and widely adopted blockchain network.
However, it is important to note that this choice of using \textsc{Polygon} as a layer 2 chain is mainly for the purpose of producing a working prototype, and further optimization and exploration with different chains is left to future work.

To implement our adaptation, we deployed a local \textsc{Polygon} chain using \textsc{Polygon Edge}\footnote{\url{https://github.com/0xPolygon/polygon-edge}: v0.7.3-beta1}.
We used \textsc{HardHat}\footnote{\url{https://github.com/NomicFoundation/hardhat}: v2.13.0}, a library that enables compilation, testing, and deployment of smart contracts, to compile and deploy ours.
Additionally, we deployed a local \textsc{Amazon S3} instance using \textsc{LocalStack}\footnote{\url{https://github.com/localstack}: v1.4.1.dev} to store the large amount of trace files.
While the use of \textsc{S3} as a storage solution for the traces may pose a potential security risk, it is important to note that it only serves as a temporary storage solution.
The security of the protocol is ensured through the transmission of the hash provided by the user on the blockchain, and the storage of the final certificate, which is also stored on-chain.
However, for future optimization, the storage system can be replaced with a more decentralized and secure solution, such as \ac{ipfs}.
This would provide a fully distributed on-chain system, ensuring the highest level of security and immutability for the stored data.
Nonetheless, this is left as an optimization for future work on the system.

Furthermore, we use a state-of-practice hash function, specifically \emph{SHA-256}, for the hashing throughout the application. In order to do so we use the \emph{hashlib} library provided by \textsc{Python}.

Finally, each of the \emph{workers} is computing the replay of the individual traces in a \textsc{Python} client using our \ac{mi} and the \textsc{web3.py} library\footnote{\url{https://github.com/ethereum/web3.py}: v6.0.0} to interact with the smart contract.

\begin{table}[tb]
	\centering
  \caption{Used symbols}
  \label{tab:symbol_recap}
  \resizebox{0.95\columnwidth}{!}{
    \sisetup{table-format=1.4}
\rowcolors{2}{}{gray!10}
\begin{tabular}{
    l l 
}

\hiderowcolors
\toprule
  \textbf{Symbol} & \textbf{Description}\\\hline
\midrule
\showrowcolors
$S$             & LIFO stack \\ 
$M$             & Program's memory \\ 
$C$             & Sequence IDs of last executed expressions for open scopes \\ 
$i$             & Index for the current position in $C$ being evaluated \\ 
$\mathcal{I}$             & Program input \\ 
$\mathcal{O}$             & Program output or a specific trace \\ 
$\mathcal{P}$             & Evaluable Mona language derivation \\ 
$\mathcal{T}$             & Set of traces \\ 
$\hat{\mathcal{T}}$             & Set of traces without outputs\\ 
$\mathT_i,\hat{\mathT_i}$           & A specific trace with or without outputs\\ 
$\mathcal{W}$             & Workers \\ 
$h$             & Hashing function \\ 
$\mathcal{H}$             & Execution hash \\ 
$h_n$           & Hash of input ($I$) or output ($O$) \\ 
$V$             & Vertices \\ 
$E$             & Edges \\ 
$v_i$           & A specific vertex \\ 
\bottomrule
\end{tabular}

  }
\end{table}

\section{Experiments}
\label{sec:experiments}

To test the feasibility of the proposed programming language (see \cref{sec:interpreter}) and the on-chain protocol (see \cref{sec:protocol}), we organized the experiments into two parts.
In the context of our study, we formulated the following research questions:
\smallskip

\begin{reqs}
\item [\req{1}] Can program executions be segmented into a collection of traces that can be re-executed to reproduce the original result and what impact does this have?
\end{reqs}

We evaluate the feasibility of the proposed programming language interpreter \ac{mi} described in \cref{sec:interpreter} by assessing its ability to produce correct and consistent output results when re-executed from a specific snapshot.
Additionally, we measure the performance overhead associated with using the proposed programming language, providing further insight into the feasibility of the approach.
By addressing these questions, we aim to provide a comprehensive assessment of the proposed programming language in terms of its ability to produce correct and consistent results, as well as its performance and scalability.
\smallskip

\begin{reqs}
	\item [\req{2}] What impact does our proposed approach have on reducing the number of executed expressions, and how does it fare in terms of robustness against malicious acts?
\end{reqs}

To assess the feasibility and efficiency of our proposed approach, we conduct an analysis to measure the number of executed expressions across various scenarios. 
These scenarios include ideal conditions (\schap), where the system operates as intended, as well as scenarios deliberately designed to mimic malicious acts. 
By quantifying the executed expressions in these different contexts, our goal is to evaluate the efficiency and resilience of our approach. 
Furthermore, all scenarios will be compared to the baseline of re-executing the full program multiple times, also referred to as naive re-execution.
We aim to provide a comprehensive evaluation of the proposed on-chain protocol and its ability to securely execute programs in a decentralized environment.

\begin{reqs}
	\item [\req{3}] How does an informed step size affect the performance of longer-running benchmark problems?
\end{reqs}
This research question investigates how an informed step size influences gas costs, certification time, and executed expressions for longer-running benchmark problems. Additionally, we aim to understand the conditions when the overheads are outweighed by the performance gains, offering a deeper understanding of its benefits.
Investigating this relationship can provide insights into choosing an appropriate step size and understanding the trade-offs between gas costs, certification time, and computational efficiency.
The informed step size is defined as the total number of expressions divided by the number of available workers, ensuring that each worker is assigned at least one trace.
This method aims to optimize resource utilization and reduce protocol-related costs. 
To evaluate this, we simulate extended benchmark scenarios by applying scaling multipliers to the existing benchmark problems.
  These results are then compared against two alternatives: the non-informed step size approach outlined in \req{2} and the baseline method described in \req{2}.

\subsection{\req{1} -- Program Segmentation}\label{subsec:rq1Exp}
To answer this question, we split the program's execution into a series of traces and re-execute them in order to reproduce the original result.
Each trace must be capable of reproducing the next state of the program and, when played in sequence, should ultimately produce the same result as running the program without splitting.
In other words, each trace $\mathT_i$ must be connected to the next $\mathT_{i+1}$ until the final trace is reached in order to ensure that the program's execution is correctly reproduced.

To evaluate the feasibility and effectiveness of the proposed trace-based certification approach for program execution, we defined the following benchmark problems:
\begin{itemize}
	\item \bpfib: We compute fibonacci, which serves as a poignant example of recursive algorithms. Additionally, it highlights the importance of optimizing recursive algorithms to avoid inefficiencies (iterative $O(n)$ vs. recursive $O(2^n)$). We compute fibonacci of 17, resulting in 69757 executed expressions.
	\item \bpfibi: A variant implementation of \bpfib that uses an iterative approach. In this case we use 98 as input resulting in 99934 executed expression.
	\item \bpmer: We evaluate merge sort on a vector of size 142 using a worst-case scenario ($O(n log n)$), yielding 99856 executed expressions. 
	\item \bpmat: We perform matrix multiplication on two matrices of dimensions $11 \times 11$ to produce matrix $C=A*B$, resulting in 86781 executed expressions. This benchmark follows established conventions \cite{parno_pinocchio_2013, zhang_transparent_2019, xie_libra_2019, wahby_doublyefficient_2018} and illustrates a time complexity of $O(n^3)$.
	\item \bpspf: We compute the shortest path using the Floyd-Warshall ($O(n^3)$) algorithm on a $13 \times 13$ matrix, resulting in 99619 executed expressions. which is used for network routing and matrix inversion, making it a common benchmark in verifiable computation schemes~\cite{parno_pinocchio_2013}.
	\item \bplaz: We employ the classic Lanczos resampling~\cite{turkowski_filters_1990} algorithm, adopted by various approaches~\cite{zhang_transparent_2019, xie_libra_2019, wahby_doublyefficient_2018}, to generate a low-resolution image from a high-resolution image. With a varying time complexity between $O(n)$ and $O(n^2)$. Using a $5 \times 5$ pixel image as input yields 76128 executed expressions.
\end{itemize}

First, these problems are well-known and widely used in the field of computer science and programming.
This ensures that our results can be compared with existing solutions and evaluated against established benchmarks.
Second, problems have varying levels of complexity and computational requirements, allowing us to evaluate the effectiveness of the proposed trace-based certification approach across a range of problem types and sizes. By testing our approach on problems with varying computational requirements, we can gain insights into its scalability and effectiveness across a range of problem types.

For each problem, we conducted a preliminary investigation to select the largest input that would lead to the evaluation of no more than \num{100000} expressions. Thus, allowing us to compare the results fairly and judge the validity of each problem based on a definitive result.

To evaluate the impact of different snapshot intervals on the performance of the trace-based certification approach, we compared the output of running the program without snapshots to replaying the snapshots at different segmentation steps (\numlist{1; 10; 100; 1000; 10000}).
We also averaged the execution time over \num{30} runs to provide a clearer view of how the overhead of the snapshots impacted the runtime.

\subsection{\req{2} -- Certification Protocol}

To examine the viability of our proposed on-chain protocol for program certification, we leverage the blockchain as a trusted, immutable, and distributed system. We assess the protocol's efficacy by executing all program traces in an unordered manner, with each worker independently determining the result of a given trace (without its output) $\hat{\mathT}$.

The protocol is engineered to certify tasks only when every provided result can be combined into a chain of traces that, when hashed, produce the hash of the target sequence. This allows us to gauge the effectiveness of the protocol by verifying whether a certificate was produced or not. 
Furthermore, we assess the efficiency of the protocol by measuring the number of executed expressions and compare it against the baseline, providing a comprehensive evaluation of its performance. 

Additionally, we record the gas costs~\cite{wood_ethereum_2017} of certifying executions on the chosen blockchain system to gain insights into the viability of running our approach on the blockchain.

For the baseline comparison, we used the \mona language along with a straightforward smart contract. Although alternative programming languages may offer faster execution, they currently lack the halt-and-resume functionality provided by our approach. 
Comparing various languages (e.g., C versus Java) would inherently result in varying execution speeds, thus, our focus is on developing a reliable certification mechanism rather than focusing on execution speed.
Our focus is on the number of executed expressions, as this remains constant regardless of parallel execution. 
In the baseline setup, each worker re-executes the entire program and votes on the correctness of the output, with certification requiring a majority vote, similar to our proposed method.
If certification fails, re-execution is performed up to three times. 
Each baseline experiment was repeated \num{30} times under conditions identical to our approach to ensure a consistent comparison.

For the assessment of our proposed on-chain protocol, we opted for the same benchmark problems as for \req{1}. Additionally, we have defined four distinct scenarios for program evaluation, namely:
\begin{itemize}
	\item \schap case, where no malicious actors participate;
	\item \sclaz case, where one or more of the workers produces an incorrect result;
	\item \scmal case, where the user attempts to certify an erroneous execution.
	\item \scera case, where another program with equivalent register values is submitted to exploit a different execution maliciously.
\end{itemize}

We conducted \num{30} runs for each of these cases to ensure statistical relevance. However, we only generated an \scera example for \bpfib to showcase the protocol's capability to handle such scenarios. 
Additionally, we measured the executed expressions in each scenario to evaluate the protocol's performance.
Based on the outcomes of preliminary experiments and feasibility analysis, we selected step sizes of \numlist{100; 1000} for protocol evaluation. These step sizes strike a balance between accuracy and performance overhead, enabling us to gain insights into the scalability of our proposed approach. 
Larger step sizes decrease the number of generated traces and computational overhead for each worker but lead to a coarser approximation of the original program execution. Conversely, smaller step sizes offer a more precise approximation of the program execution but increase the number of generated traces and the computational overhead for each worker.
For all experiments, we employed \num{20} workers in parallel to retrieve tasks from the smart contract and replay the assigned traces. 
Additionally, we utilized \num{3} workers to verify the proposed results and assess them for conflicts.

\paragraph{Simulation cases description}
In the \schap scenario, we assume that none of the participating actors act maliciously and measure the efficiency of our approach by comparing the number of executed expressions against the baseline, providing insights into the performance impact of trace-based program certification on a blockchain-based platform.
Additionally, we measure the time required for certification to gain insights into the overhead produced by different step sizes and indirectly its impact on the blockchain infrastructure overhead.

In the \sclaz scenario, we introduce up to 40\% malicious workers who randomly select one of the other available traces and votes for it instead of replaying the assigned trace.
Our approach should still be able to produce a certificate and detect the conflicts introduced by the lazy workers. However, we expect the number of executed expressions for this scenario to be higher than the \schap scenario due to the additional computational overhead required to detect and resolve the conflicts.

In the \scmal scenario, we assume that the user produces all the required snapshots to replay the program but intentionally provides a different result than the actual output that is supposed to be certified. For example, the user may provide \code{fibonacci(17) = 5} instead of the correct result \code{fibonacci(17) = 1597}. We anticipate the number of executed expressions for this scenario to be similar to the \schap scenario since the introduced mismatch is simple and does not require significant additional computation to detect and resolve.

In the \scera scenario, the user furnishes an alternate implementation of the algorithm, generating identical register values as the original version. More precisely, we present an iterative implementation for \bpfib in lieu of the recursive approach. We expect the protocol to discern this variation and terminate the certification process. Additionally, we anticipate the number of executed expressions for this scenario to closely resemble those of the \scmal scenario.
  \subsection{\req{3} -- Informed step size}\label{subsec:exp:rq3}
To investigate this question, we assess the impact of using an informed step size on tasks with increased computational demands. 
Benchmark problems are scaled using scaling multipliers \num{1}, \num{10}, \num{100}, and \num{1000}, which proportionally increase the workload by raising the number of executed expressions. 
However, this scaling is approximate because additional expressions are not always generated when defining functions or referencing existing variables. Details of these variations are provided in \cref{tab:rq3_bench}.

The informed step size is calculated as $\frac{\text{expressions}}{|W|}$, where $|W|$ is the number of workers, ensuring each worker processes at least one trace. 
To evaluate its effect, we compare the performance of the informed step size against the naive re-execution baseline from \req{2}, examining differences in time, gas usage, and executed expressions. 
The results from multiplier \num{1} are also compared to the outcomes of \req{2} without an informed step size to isolate its direct impact.

For these experiments, the parameters from \req{2} are reused with modifications: the step size is adjusted, and the number of reruns is reduced to three (from 30) to accommodate time and resource constraints. 
Although the informed step size is not necessarily the most efficient solution, this analysis provides useful insights into its trade-offs and informs future optimization approaches.
\begin{table}[tb]
	\centering
  \caption{\req{3} Scaled Benchmark Problems}
  \label{tab:rq3_bench}
  \resizebox{0.95\columnwidth}{!}{
	\sisetup{table-format=1.4}
\rowcolors{2}{}{gray!10}
\begin{tabular}{
    l S[table-format=5] S[table-format=6] S[table-format=7] S[table-format=8] 
}

\hiderowcolors
\toprule
  {\multirow{2}[2]{*}{\textbf{Program}}} & \multicolumn{4}{c}{\textbf{Exec. Exprs}}  \\\cmidrule{2-5}
  
  & \textbf{\num{1}} & \textbf{\num{10}} & {\textbf{\num{100}}} & {\textbf{\num{1000}}}\\
\midrule
\showrowcolors
  \bpfib & 69757 & 697552 & 6975502 & 69755002 \\
  \makecell[l]{\textsc{Fibonacci}\\\textsc{Iterative}} & 99934 & 999304 & 9993004 & 99930004\\
  \bplaz & 76128 & 761226 & 7612206 & 76122006\\
  \makecell[l]{\textsc{Matrix}\\\textsc{Multiplication}} & 86781 & 866550 & 8664240 & 86641140 \\
  \bpmer & 99856 & 997219 & 9970849 & 99707149 \\
  \makecell[l]{\textsc{Shortest}\\\textsc{PathFirst}} & 99619 & 996163 & 9961603 & 99616003 \\
\bottomrule
\end{tabular}
 
  }
\end{table}

\subsection{Execution Setup}

To ensure the consistency of the experiments, we ran all experiments on the same machine with the same specifications, \ie, Xeon Gold 6126 with \num{32} vCPUs and \qty{256}{\giga\byte} of RAM.
The virtual machine was hosted on a cloud computing platform with dedicated resources, ensuring that there were no performance fluctuations due to shared hardware or resource contention.
Additionally, the machine was running Ubuntu 22.04 LTS, and all experiments were conducted using \textsc{Python} v3.10.6.

\section{Results}
\label{sec:results}

This section presents the results obtained from the experiments conducted in \cref{sec:experiments}.
For each research question, we provide detailed insights on the proposed approach and the corresponding validation method used for the experiments.
In comparing observations, the Kruskal-Wallis test revealed overall group differences, and post hoc Mann-Whitney U rank tests confirmed pairwise distinctions, all with p-values $\leq 0.05$. Significance at $\alpha = 0.05$ level was established. Effect sizes were computed using the Rank-Biserial Correlation coefficient (r), yielding an r value of 1, signifying a substantial and statistically significant difference between the samples' distributions.

\subsection{\req{1} -- Program Segmentation}
\begin{table}[tb]
	\centering
  \caption{\req{1} results -- Average runtime in seconds for \num{30} iterations}
  \label{tab:rq1_results}
  \resizebox{0.95\columnwidth}{!}{
	\rowcolors{2}{}{gray!10}
\begin{tabular}{
    S[table-format=5] S[table-format=3.3] S[table-format=3.3] S[table-format=3.3] S[table-format=5] S[table-format=3.3]
}

\hiderowcolors
\toprule

{\multirow{2}[2]{*}{\textbf{Step Size}}} & \multicolumn{3}{c}{\textbf{Avg. Time (\unit{\second})}} & \multicolumn{2}{c}{\textbf{Space requirements}} \\
\cmidrule(lr){2-4}
\cmidrule(lr){5-6}
& {\textbf{Execution}} & {\textbf{Recording}} & {\textbf{Replay}} & {\textbf{\makecell[r]{Num.\\ Snapshots}}} & {\textbf{\makecell[r]{Avg.\\ size (KB)}}} \\ 

\midrule
\showrowcolors

\multicolumn{6}{c}{\bpfibi}\\
1 & 0.223 & 148.750 & 227.446 & 99935 & 4.008      \\
10 & 0.223 & 14.895 & 22.661  & 9995  & 4.008      \\
100 & 0.223 & 1.720 & 2.401   & 1001  & 4.010      \\
1000 & 0.223 & 0.391 & 0.442  & 101   & 3.988   \\
10000 & 0.223 & 0.256 & 0.265 & 11    & 3.910   \\
\midrule

\multicolumn{6}{c}{\bpfib}\\
1 & 0.180 & 60.519 & 127.933  & 69758& 3.667       \\
10 & 0.180 & 6.280 & 12.703   & 6977 & 3.667       \\
100 & 0.180 & 0.819 & 1.444   & 699  & 3.657    \\
1000 & 0.180 & 0.257 & 0.313  & 71   & 3.597    \\
10000 & 0.180 & 0.197 & 0.206 & 8    & 2.882    \\
\midrule

\multicolumn{6}{c}{\bpmer}\\
1       & 0.255 & 545.916 & 592.353 & 99857  & 21.221     \\
10      & 0.255 & 54.520 & 56.779   & 9987   & 21.220    \\
100     & 0.255 & 5.459 & 5.429     & 1000   & 21.215    \\
1000    & 0.255 & 0.791 & 0.762     & 101    & 20.836    \\
10000   & 0.255 & 0.305 & 0.310     & 11     & 10.623 \\
\midrule

\multicolumn{6}{c}{\bpmat}\\
1       &0.224 & 327.370 & 402.173  & 86782 & 14.121  \\
10      &0.224 & 32.667 & 38.833    & 8680  & 14.118  \\
100     &0.224 & 3.298 & 3.777      & 869   & 14.103  \\
1000    &0.224 & 0.549 & 0.571      & 88    & 13.851 \\
10000   &0.224 & 0.268 & 0.276      & 10    & 11.976 \\
\midrule

\multicolumn{6}{c}{\bpspf} \\
1       &0.265 & 943.317 & 808.596   &  99620 & 20.394 \\
10      &0.265 & 94.804 & 81.469    &  9963   & 20.394   \\
100     &0.265 & 9.647 & 8.200      &  998    & 20.368   \\
1000    &0.265 & 1.102 & 0.837      &  101    & 20.143   \\
10000   &0.265 & 0.361 & 0.342      &  11     & 17.955   \\
\midrule

\multicolumn{6}{c}{\bplaz} \\
1       & 0.219 & 594.447 & 588.258  & 76129 & 14.715 \\
10      & 0.219 & 59.865 & 57.526    & 7614  & 14.714  \\
100     & 0.219 & 6.146 & 5.888      & 763   & 14.699 \\
1000    & 0.219 & 0.804 & 0.762      & 78    & 14.449\\
10000   & 0.219 & 0.281 & 0.277      & 9     & 12.585\\
\bottomrule
\end{tabular}

  }
\end{table}
Our experiments indicate that the \acf{mi} effectively records and replays accurate results across all step sizes.
To verify that each trace $\mathT_i$ produces the correct output $\mathO_i$ we ran all the traces $\mathT$ in sequence and compared the produced output to the oracle, \ie, the hash of the next trace $\mathT_{i+1}$.
Additionally, our results suggest that the runtime for execution, recording, and replay decreases as the step size increases for all programs.
However, a trade-off exists between trust (low step size) and performance (high step size).
While a step size of one provides accurate and trustworthy results, it is impractical due to its longer runtime.
Conversely, a higher step size improves performance but may affect the trustworthiness of the results, as shown in \cref{tab:rq1_results}.
Although full trust can only be achieved with a step size of one, further research and experiments are required to determine the actual impact on performance and trust.
Notably, the step size also affects space requirements: the memory needed per snapshot, on average, corresponds to the memory usage of the program under evaluation until the subsequent snapshot. 
In our experiments, the average space requirement per snapshot ranged from 4 to 21 KB as shown in \cref{tab:rq1_results}. 
However, a decrease in step size leads to a higher number of snapshots, thereby increasing the total memory demand. 
This indicates that the step size is directly related to the performance and space consumption.
Further research is necessary to thoroughly assess the balance between performance, trustworthiness, and memory usage.

Our prototype language was able to reproduce accurate results for the given benchmark problems, as shown in \cref{tab:rq1_results}.
However, some of the benchmark problems exhibited unexpected results, as the replay time was higher than the recording time.
This deviation is the result of the large tail recursion promoted by these programs, which caused every replay to rebuild large proportionally deep evaluation trees.
Further research is necessary to identify potential optimizations for the \ac{mi} and to investigate how these optimizations can improve the replay time of programs with large tail recursion.

\begin{custombox}[\req{1} -- In summary]
	The experiments demonstrated that the \ac{mi} prototype is effective in accurately recording and replaying program executions for all step sizes and benchmark problems.
  The trade-off between trust and performance was observed, with lower step sizes providing higher trust but lower performance, and higher step sizes offering better performance at the cost of lower trust.
  Additionally, the experiments revealed that lower step sizes result in increased space requirements.
  This reflects a direct relationship among step size, performance, and space requirements, underscoring the intricate balance that must be managed between these factors.
\end{custombox}

\subsection{\req{2} -- Certification Protocol}\label{subsec:results:rq2}

\begin{table*}[tb]
	\caption{\req{2} results -- Average runtime, in seconds, and number of executed expressions, over \num{30} runs}
	\label{tab:rq2_results}
	\centering
  \resizebox{\linewidth}{!}{
    \sisetup{table-format=1.4}
\rowcolors{2}{}{gray!10}
\begin{tabular}{
    l l S[table-format=3.3] S[table-format=3.3] S[table-format=3.3] S[table-format=6.3] S[table-format=6.3] S[table-format=6.3] S[table-format=6.3] S[table-format=6.3] S[table-format=6.3] c c c 
}

\hiderowcolors
\toprule

  {\multirow{3}[3]{*}{\textbf{Program}}} & {\multirow{3}[3]{*}{\textbf{Scenario}}} & \multicolumn{3}{c}{\textbf{Avg. Cert. Time (\unit{\second})}} & \multicolumn{3}{c}{\textbf{Gas costs (Mil.)}} & \multicolumn{3}{c}{\textbf{Exec. Exprs}} & \multicolumn{3}{c}{\textbf{Error rates}} \\
  \cmidrule(lr){3-5} \cmidrule(lr){6-8} \cmidrule(lr){9-11} \cmidrule(lr){12-14}
  & & \multicolumn{3}{c}{\textbf{Step Size}} & \multicolumn{3}{c}{\textbf{Step Size}} & \multicolumn{3}{c}{\textbf{Step Size}}& \multicolumn{3}{c}{\textbf{FP/FN}}\\
  \cmidrule(lr){3-5} \cmidrule(lr){6-8} \cmidrule(lr){9-11} \cmidrule(lr){12-14}
  & & {\textbf{\textsc{BaseLine}}} & {\textbf{\num{100}}} & {\textbf{\num{1000}}} & {\textbf{\textsc{BaseLine}}} & {\textbf{\num{100}}} & {\textbf{\num{1000}}} & {\textbf{\textsc{BaseLine}}} & {\textbf{\num{100}}} & {\textbf{\num{1000}}} & {\textbf{\textsc{BaseLine}}} & {\textbf{\num{100}}} & {\textbf{\num{1000}}} \\

\midrule
\showrowcolors
  \bpfib  & \makecell[l]{\textsc{Equivalent}\\\textsc{RegistersAttack}\@\xspace} \cellcolor{gray!10}       & 12.716 & 513.174   & 73.652    & 4.202   & 373.748   & 39.329   & 71360.0     & 100.0     & 1000.0       & 1.0/0.0        & 0.0/0.0   & 0.0/0.0    \\
          & \schap        & 15.555 & 514.896   & 73.579    & 3.722   & 476.797   & 49.829   & 1395140.0   & 69757.0   & 76732.7      & 0.0/0.0     & 0.0/0.0   & 0.0/0.0    \\     
          & \sclaz 10\%   & 15.387 & 654.382   & 135.912   & 3.616   & 598.401   & 69.822   & 1255626.0   & 72663.667 & 95600.3      & 0.0/0.0     & 0.0/0.0   & 0.0/0.0    \\
          & \sclaz 20\%   & 15.116 & 702.067   & 174.426   & 3.515   & 617.385   & 79.706   & 1116112.0   & 74730.334 & 128793.133   & 0.0/0.1     & 0.0/0.0   & 0.0/0.0    \\      
          & \sclaz 30\%   & 14.989 & 823.23    & 435.668   & 7.176   & 617.69    & 117.174  & 1953196.0   & 78107.0   & 262824.9     & 0.0/0.033   & 0.0/0.0   & 0.0/0.0    \\      
          & \sclaz 40\%   & 14.3   & 1057.348  & 1274.64   & 7.07    & 593.071   & 244.263  & 1674168.0   & 85274.833 & 659872.834   & 0.0/0.567   & 0.0/0.0   & 0.0/0.0    \\      
          & \scmal        & 17.011 & 515.634   & 73.639    & 3.318   & 471.865   & 49.089   & 1395140.0   & 69757.0   & 76732.7      & 0.0/0.0     & 0.0/0.0   & 0.0/0.0    \\      
\midrule
  \makecell[l]{\textsc{Fibonacci}\\\textsc{Iterative}} & \schap        & 18.515& 729.395   & 87.593    & 4.12  & 682.689 & 69.866  & 1998680.0 & 99934.0     & 99934.0      & 0.0/0.0          & 0.0/0.0  & 0.0/0.0 \\
          & \sclaz 10\%   & 16.823& 954.152   & 221.272   & 3.942 & 866.128 & 91.811  & 3370950.334 & 102170.667  & 125100.667   & 0.0/0.0        & 0.0/0.0  & 0.0/0.0 \\
          & \sclaz 20\%   & 16.403& 965.92    & 202.965   & 3.968 & 887.046 & 105.304 & 3277958.6 & 90196.133   & 154156.334   & 0.0/0.567    & 0.0/0.0  & 0.0/0.0 \\
          & \sclaz 30\%   & 15.017& 1066.666  & 428.06    & 4.003 & 814.456 & 132.508 & 2680154.6 & 106184.0    & 260056.666   & 0.0/0.733    & 0.0/0.0  & 0.0/0.0 \\
          & \sclaz 40\%   & 14.742& 1411.262  & 1171.964  & 3.879 & 857.327 & 244.888 & 2328114.467 & 110517.4    & 623560.533   & 0.0/0.933    & 0.0/0.0  & 0.0/0.0 \\
          & \scmal        & 19.194& 730.487   & 87.741    & 3.849 & 675.636 & 68.894  & 4387217.134 & 99934.0     & 99934.0      & 0.0/0.0        & 0.0/0.0  & 0.0/0.0 \\
\midrule
  \bplaz    & \schap      & 16.117& 581.605 & 76.26    &7.976 & 521.053 & 54.467  & 1522560.0& 76128.0     & 76128.0      & 0.0/0.0   & 0.0/0.0 & 0.0/0.0 \\ 
            & \sclaz 10\% & 16.546& 721.855 & 141.194  &7.723 & 640.154 & 74.879  & 1370304.0& 78514.667   & 115495.466   & 0.0/0.0   & 0.0/0.0 & 0.0/0.0 \\ 
            & \sclaz 20\% & 16.674& 779.922 & 176.752  &7.804 & 639.503 & 83.339  & 1218048.0& 80578.0     & 129516.0     & 0.0/0.0   & 0.0/0.0 & 0.0/0.0 \\ 
            & \sclaz 30\% & 15.078& 938.292 & 389.54   &15.666& 622.745 & 109.196 & 2131584.0& 84032.266   & 222675.734   & 0.0/0.166 & 0.0/0.0 & 0.0/0.0 \\ 
            & \sclaz 40\% & 15.064& 1180.06 & 1144.937 &15.516& 649.454 & 220.22  & 1827072.0& 92321.333   & 563151.2     & 0.0/0.266 & 0.0/0.0 & 0.0/0.0 \\ 
            & \scmal      & 19.31 & 579.247 & 78.147   &7.442 & 515.845 & 53.64   & 1522560.0& 76128.0     & 76128.0      & 0.0/0.0   & 0.0/0.0 & 0.0/0.0 \\ 
\midrule
  \makecell[l]{\textsc{Matrix}\\\textsc{Multiplication}} & \schap       &15.807& 651.147    & 90.916    &7.037 & 593.492 & 61.552  &1735620.0& 86781.0     & 86781.0    & 0.0/0.0   & 0.0/0.0 & 0.0/0.0 \\ 
          & \sclaz 10\%  &16.087& 779.441    & 142.687   &6.77  & 804.691 & 79.721  &1562058.0& 88994.5     & 109533.067 & 0.0/0.0   & 0.0/0.0 & 0.0/0.0 \\
          & \sclaz 20\%  &16.49 & 887.788    & 214.627   &6.872 & 678.446 & 94.805  &1388496.0& 90607.667   & 148134.033 & 0.0/0.033 & 0.0/0.0 & 0.0/0.0 \\
          & \sclaz 30\%  &15.116& 1084.062   & 433.089   &13.745& 723.366 & 123.869 &2429868.0& 98074.333   & 255178.533 & 0.0/0.05  & 0.0/0.0 & 0.0/0.0 \\
          & \sclaz 40\%  &14.972& 1210.444   & 2175.75   &13.609& 724.327 & 330.615 &2082744.0& 100784.367  & 949897.0   & 0.0/0.366 & 0.0/0.0 & 0.0/0.0 \\
          & \scmal       &19.287& 650.831    & 91.947    &5.088 & 587.359 & 60.662  &1735620.0& 86781.0     & 86781.0    & 0.0/0.0   & 0.0/0.0 & 0.0/0.0 \\
\midrule
  \bpmer  & \schap      &16.38 & 736.882  & 96.897    &6.532 & 682.675 & 70.084  &1997120.0  & 99856.0     & 99856.0      & 0.0/0.0   & 0.0/0.0 & 0.0/0.0 \\ 
          & \sclaz 10\% &17.667& 962.512  & 175.203   &6.361 & 980.638 & 92.182  &1797408.0  & 102279.333  & 123156.0     & 0.0/0.0   & 0.0/0.0 & 0.0/0.0 \\
          & \sclaz 20\% &16.524& 964.519  & 195.528   &6.378 & 818.057 & 106.097 &1597696.0  & 103719.333  & 155817.867   & 0.0/0.0   & 0.0/0.0 & 0.0/0.0 \\
          & \sclaz 30\% &15.305& 1122.631 & 453.675   &12.794& 821.501 & 138.285 &2795968.0  & 105909.333  & 276141.066   & 0.0/0.1   & 0.0/0.0 & 0.0/0.0 \\
          & \sclaz 40\% &15.022& 1552.131 & 780.445   &12.632& 841.294 & 193.359 &2396544.0  & 117894.534  & 441306.933   & 0.0/0.316 & 0.0/0.0 & 0.0/0.0 \\
          & \scmal      &19.664& 739.686  & 92.292    &5.406 & 675.584 & 69.112  &2020419.734& 99856.0     & 99856.0      & 0.0/0.0   & 0.0/0.0 & 0.0/0.0 \\
\midrule
  \makecell[l]{\textsc{Shortest}\\\textsc{PathFirst}}  & \schap      &18.028& 762.622  & 97.907    &8.252 & 681.703  & 70.002    &1992380.0& 99619.0      & 99619.0      & 0.0/0.0   & 0.0/0.0 & 0.0/0.0 \\ 
          & \sclaz 10\% &17.472& 894.811  & 168.344   &8.182 & 816.025  & 90.457    &1793142.0& 101725.667   & 119542.8     & 0.0/0.0   & 0.0/0.0 & 0.0/0.0 \\
          & \sclaz 20\% &17.022& 953.828  & 233.937   &8.184 & 830.349  & 95.614    &1593904.0& 103682.333   & 148934.867   & 0.0/0.0   & 0.0/0.0 & 0.0/0.0 \\
          & \sclaz 30\% &16.436& 1115.191 & 711.422   &16.456& 779.423  & 156.214   &2789332.0& 106282.333   & 355760.266   & 0.0/0.083 & 0.0/0.0 & 0.0/0.0 \\
          & \sclaz 40\% &15.032& 1501.588 & 1265.742  &16.145& 823.51   & 238.196   &2390856.0& 115730.4     & 609861.733   & 0.0/0.366 & 0.0/0.0 & 0.0/0.0 \\
          & \scmal      &21.084& 773.87   & 99.478    &7.071 & 674.618  & 69.03     &2042189.5& 99619.0      & 99619.0      & 0.0/0.0   & 0.0/0.0 & 0.0/0.0 \\
\bottomrule
\end{tabular}

  }
\end{table*}

The experiments have demonstrated the feasibility of running the \ac{occp} on a Layer 2 blockchain, specifically \textsc{Polygon}, as presented in \cref{tab:rq2_results}. 

When comparing the trade-off between step sizes of \num{100} and \num{1000} in the \sclaz scenario, there is an increase in the number of executed expressions in the latter, and conversely, a decrease in the required certification time. 
This is due to the fact that the increased step size reduces the communication overhead on the blockchain platform but requires more expressions for any given trace to be re-executed. 

When comparing the results of the \sclaz scenario experiments, we observe that our approach consistently results in fewer re-executed expression than the baseline of the respective scenario.
The increase in executed expressions observed in the \sclaz scenario, compared to the \schap scenario, is due to the need to re-execute traces that were
assigned to malicious workers. 
This difference arises because malicious workers cast votes without executing the traces, which requires additional re-execution of the involved traces to resolve the resulting conflicts. 
On average, a malicious worker incurs an additional overhead of expressions required for executing one trace, whereas full re-execution would double the executed expressions.
Furthermore, a higher gas cost is associated with a smaller step size, these costs can be attributed to the higher communication requirements.
However, these costs can be controlled through step size adjustments. Specifically, a larger step size not only improves performance by reducing communication overhead but also helps mitigate gas consumption.
Future work should therefore focus on finding a trade-off between smart contract optimization and efficient gas usage.

Our approach reliably handles up to 40\% malicious workers, demonstrating its robustness in detecting and addressing malicious behavior. 
Unlike the baseline approach, which fails to detect subtle code modifications unless they affect the final result, our method consistently identifies such manipulations as demonstrated by the \scera scenario. 
The integration of our protocol's sequence hash verification and trace-based certification not only addresses the limitations of the naive approach but also enables reliable execution certification by detecting input and output manipulations as well as code modifications, resulting in a more robust solution.
However, the increased reliability comes with a trade-off of higher resource costs. Notably, when the percentage of malicious workers reaches 40\%, the overhead introduced becomes significant. 
Despite this, the system remains effective in certifying tasks even in the presence of additional malicious workers.
Further research is required to gauge the impact of multiple malicious workers in combination with malicious users (see \cref{sec:experiments:threats}).

When comparing the \scmal results to the \schap results regarding executed expressions, we noticed that the executed expressions were the same, indicating that the number of executed expressions was not impacted by whether the execution was successful or rejected. 
This lack of difference was anticipated as traces for a full program are provided in both scenarios; however, no quorum is reached in the \scmal scenario. 
Thus, the workers need to compute each trace individually as in the \schap scenario and finally reach the conclusion that the execution does not match the provided \ac{eh}.

We wish to stress that a rejected execution is not necessarily due to malicious acts, and a false negative has less weight than a successful certification. 
To restate, our approach has higher resilience to false positives than false negatives. 
Overall, the results demonstrate that our approach is effective in detecting and resolving conflicts in a distributed computation setting and can achieve reasonable performance and scalability on a Layer 2 blockchain.

The average communication time between the workers and the \smartcontract is \num{6.353} seconds. 
When analyzing the impact of different scenarios and step sizes, we observe that an increase in the number of traces leads to a corresponding increase in communication overhead. 
Specifically, in the \schap scenario, the average communication time increases 10-fold when the step size is increased from \num{100} to \num{1000}.
For the \sclaz scenarios, the communication overhead varies due to the additional interactions introduced by conflicts. The observed increases in communication for \sclaz of 10\%, 20\%, 30\%, and 40\% of malicious workers were \num{8}-fold, \num{5}-fold, \num{3}-fold, and \num{2}-fold, respectively.
In the baseline approach, the average number of communications is fixed at \num{20} for \schap scenario, while the average communication time between workers and the \smartcontract remains consistent with our proposed approach. 
However, the average overhead introduced by our method, compared to the baseline, is \num{4}-fold for a step size of \num{1000} and nearly \num{44}-fold for a step size of \num{100}.
This overhead can be reduced by adjusting the step size, which will result in fewer communication events.

\begin{custombox}[\req{2} -- In summary]
  Our experiments have demonstrated that the proposed \ac{occp} is a feasible approach where the overhead of time and resources are dependent on the chosen step size, as observed in \req{1}.
However, in this initial research, we observe performance degradation due to implementation challenges, such as communication overhead from blockchain interactions, which need to be optimized in future work. 
In addition, our approach can handle all proposed malicious scenarios and reliably certify tasks or reject them. 
\end{custombox}

\subsection{\req{3} -- Informed step size}\label{subsec:res:rq3}
The results show that the informed step size, compared to the non-informed variations (see \req{2}), reduces the required certification time by up to 26-fold for the step size of \num{100} and 10-fold for \num{1000}. 
Similarly, gas costs are reduced by up to 44-fold for a step size of \num{100} and 6-fold for \num{1000}. 
As expected, the number of executed expressions for the multiplier \num{1} remains approximately the same as reported in \req{2}.

Compared to the baseline, our approach consistently demonstrated performance gains for benchmarks scaled with a multiplier of \num{1000}. 
In this configuration, our method required up to 43-fold less time and 12-fold less gas consumption. 
However, these performance improvements only became apparent at or beyond the threshold multiplier of \num{1000}, indicating that performance gains outweigh communication overheads as the program size increases.
For smaller problem sizes (multipliers below \num{1000}), our informed step size approach incurred higher time and gas costs during the certification process, similar to the trends observed in \req{2}.
Nonetheless, benchmarks scaled with a multiplier of \num{100} already exhibited notable time savings of up to 9-fold, though gas costs remained higher, with up to a 3-fold increase compared to the baseline. 
Despite these increases for smaller multipliers, our approach consistently reduced the need for re-execution of expressions across all experimental configurations.

We observe that in the baseline, the required execution time decreases as the number of malicious workers increases. This behavior is consistent with the findings in \req{2}, where malicious workers avoid executing the intended program, effectively skipping computational work. The reduction in execution time is more pronounced for larger program sizes and higher proportions of malicious workers, as the skipped workload scales with these factors.

\begin{table*}[tb]
  \centering
  \caption{\req{3} results -- Impact comparison of long-running executions.\\Average runtime, in seconds, and number of executed expressions, over \num{3} runs.}
  \label{tab:rq3_results_long}
  \resizebox{\linewidth}{!}{
    \sisetup{table-format=1.4}
\rowcolors{2}{}{gray!10}
\begin{tabular}{
    l l l S[table-format=3.3] S[table-format=3.3] S[table-format=3.3] S[table-format=3.3] S[table-format=6.3] S[table-format=6.3] S[table-format=6.3] S[table-format=6.3] S[table-format=6.3] S[table-format=6.3] S[table-format=6.3] S[table-format=6.3]
}

\hiderowcolors
\toprule
  \multirow{3}{*}{\textbf{Program}} & \multirow{3}{*}{\textbf{Scenario}} & \multirow{3}{*}{\textbf{Approach}} & \multicolumn{4}{c}{\textbf{Avg. Cert. Time (\unit{\second})}} & \multicolumn{4}{c}{\textbf{Gas costs (Mil.)}} & \multicolumn{4}{c}{\textbf{Exec. Exprs (Mil.)}} \\
  \cmidrule(lr){4-7} \cmidrule(lr){8-11} \cmidrule(lr){12-15}
  & & & \multicolumn{4}{c}{Multiplier} &\multicolumn{4}{c}{Multiplier} &\multicolumn{4}{c}{Multiplier}\\
  \cmidrule(lr){4-7} \cmidrule(lr){8-11} \cmidrule(lr){12-15}
  & & & \textbf{1} & \textbf{10} & \textbf{100} &\textbf{1000} & \textbf{1} & \textbf{10} & \textbf{100} &\textbf{1000} & \textbf{1} & \textbf{10} & \textbf{100} &\textbf{1000} \\

\midrule
\showrowcolors
 &  & \textsc{naive} & 14.941 & 28.159 & 161.941& 1476.378 & 3.869 & 4.416 & 7.099& 32.552 & 1.395 & 13.951 & 139.51& 1395.1\\
 & \multirow{-2}{*}{ \schap} & \textsc{occp} & 27.163 & 30.189 & 48.118& 238.078 & 15.536 & 15.591 & 15.529& 15.571 & 0.07 & 0.698 & 6.976& 69.755\\
 &  & \textsc{naive} & 15.19 & 33.202 & 152.703& 8222.387 & 3.7 & 5.136 & 9.596& 224.273 & 1.256 & 14.021 & 139.51& 2790.2\\
 & \multirow{-2}{*}{ \sclaz 10\%} & \textsc{occp} & 45.712 & 109.518 & 100.672& 412.286 & 24.951 & 21.163 & 22.148& 21.554 & 0.105 & 1.099 & 11.242& 122.188\\
 &  & \textsc{naive} & 17.744 & 73.7 & 614.431& 5955.659 & 3.634 & 4.572 & 8.301& 37.236 & 1.116 & 11.161 & 111.608& 1116.08\\
 & \multirow{-2}{*}{ \sclaz 20\%} & \textsc{occp} & 80.157 & 172.494 & 112.192& 732.102 & 31.18 & 25.077 & 27.127& 32.474 & 0.14 & 1.431 & 17.264& 248.444\\
 &  & \textsc{naive} & 17.563 & 65.545 & 540.228& 5212.203 & 3.557 & 5.035 & 7.85& 35.833 & 0.977 & 9.766 & 97.657& 976.57\\
 & \multirow{-2}{*}{ \sclaz 30\%} & \textsc{occp} & 90.191 & 151.625 & 202.538& 1420.791 & 34.475 & 30.153 & 36.69& 55.965 & 0.196 & 1.821 & 27.228& 473.172\\
 &  & \textsc{naive} & 15.553 & 57.552 & 463.555& 4462.853 & 3.511 & 4.054 & 8.309& 37.569 & 0.837 & 8.371 & 83.706& 837.06\\
 & \multirow{-2}{*}{ \sclaz 40\%} & \textsc{occp} & 163.163 & 271.201 & 388.817& 1870.986 & 45.516 & 54.254 & 60.565& 68.729 & 0.337 & 4.466 & 53.444& 600.939\\
 &  & \textsc{naive} & 16.936 & 54.312 & 416.188& 4019.539 & 3.274 & 5.165 & 8.186& 36.878 & 1.395 & 13.951 & 139.51& 1395.1\\
\multirow{-12}{*}{ \bpfib} & \multirow{-2}{*}{ \scmal} & \textsc{occp} & 27.12 & 30.489 & 48.497& 241.328 & 15.111 & 15.122 & 15.061& 15.101 & 0.07 & 0.698 & 6.976& 69.755\\
\midrule
 &  & \textsc{naive} & 18.443 & 63.167 & 505.242& 4680.364 & 3.879 & 5.708 & 7.047& 19.124 & 1.999 & 19.986 & 199.86& 1998.6\\
 & \multirow{-2}{*}{ \schap} & \textsc{occp} & 27.398 & 30.6 & 53.049& 283.662 & 15.593 & 15.541 & 15.572& 15.58 & 0.1 & 0.999 & 9.993& 99.93\\
 &  & \textsc{naive} & 16.969 & 33.003 & 181.573& 10026.82 & 3.815 & 5.459 & 8.615& 69.538 & 1.799 & 19.953 & 199.86& 1998.6\\
 & \multirow{-2}{*}{ \sclaz 10\%} & \textsc{occp} & 61.745 & 99.833 & 102.112& 505.118 & 26.535 & 20.945 & 22.801& 22.177 & 0.174 & 1.581 & 16.838& 195.03\\
 &  & \textsc{naive} & 19.089 & 86.409 & 755.664& 7395.015 & 3.849 & 4.855 & 6.845& 18.709 & 1.599 & 15.989 & 159.888& 1598.88\\
 & \multirow{-2}{*}{ \sclaz 20\%} & \textsc{occp} & 61.594 & 105.665 & 125.739& 807.301 & 31.855 & 24.64 & 26.822& 31.717 & 0.235 & 1.99 & 24.116& 330.602\\
 &  & \textsc{naive} & 18.231 & 77.559 & 658.887& 6452.206 & 3.913 & 5.377 & 6.557& 18.103 & 1.399 & 13.99 & 139.902& 1399.02\\
 & \multirow{-2}{*}{ \sclaz 30\%} & \textsc{occp} & 93.541 & 165.477 & 192.476& 1377.47 & 35.913 & 32.047 & 34.569& 48.167 & 0.321 & 3.018 & 36.908& 550.281\\
 &  & \textsc{naive} & 17.557 & 70.205 & 568.222& 5550.851 & 3.912 & 5.703 & 6.936& 19.013 & 1.199 & 11.992 & 119.916& 1199.16\\
 & \multirow{-2}{*}{ \sclaz 40\%} & \textsc{occp} & 142.367 & 292.566 & 410.1& 3542.724 & 43.655 & 46.044 & 63.349& 101.065 & 0.45 & 4.995 & 81.143& 1373.371\\
 &  & \textsc{naive} & 18.74 & 63.075 & 505.963& 4681.979 & 3.902 & 5.699 & 7.06& 19.137 & 1.999 & 19.986 & 199.86& 1998.6\\
\multirow{-12}{*}{ \bpfibi} & \multirow{-2}{*}{ \scmal} & \textsc{occp} & 27.219 & 30.565 & 54.117& 289.367 & 15.122 & 15.073 & 15.108& 15.112 & 0.1 & 0.999 & 9.993& 99.93\\
\midrule
 &  & \textsc{naive} & 15.225 & 29.832 & 168.942& 1494.42 & 8.088 & 8.605 & 9.927& 20.733 & 1.523 & 15.225 & 152.244& 1522.44\\
 & \multirow{-2}{*}{ \schap} & \textsc{occp} & 28.103 & 30.248 & 50.731& 242.947 & 15.543 & 15.517 & 15.534& 15.545 & 0.076 & 0.761 & 7.612& 76.122\\
 &  & \textsc{naive} & 15.166 & 33.268 & 159.649& 8633.519 & 7.808 & 9.853 & 13.339& 78.675 & 1.37 & 15.275 & 152.244& 1522.44\\
 & \multirow{-2}{*}{ \sclaz 10\%} & \textsc{occp} & 56.341 & 94.173 & 104.051& 445.491 & 25.772 & 20.861 & 22.279& 22.128 & 0.125 & 1.166 & 13.106& 146.789\\
 &  & \textsc{naive} & 19.748 & 77.709 & 642.356& 6245.005 & 8.026 & 9.132 & 11.418& 23.916 & 1.218 & 12.18 & 121.795& 1217.952\\
 & \multirow{-2}{*}{ \sclaz 20\%} & \textsc{occp} & 78.167 & 140.013 & 126.56& 610.978 & 31.687 & 25.112 & 26.326& 29.332 & 0.173 & 1.6 & 17.66& 228.747\\
 &  & \textsc{naive} & 17.566 & 69.571 & 561.562& 5456.849 & 7.724 & 9.768 & 10.957& 22.748 & 1.066 & 10.657 & 106.571& 1065.708\\
 & \multirow{-2}{*}{ \sclaz 30\%} & \textsc{occp} & 120.18 & 189.148 & 208.381& 1985.648 & 34.328 & 37.461 & 36.873& 70.365 & 0.216 & 2.914 & 31.489& 687.255\\
 &  & \textsc{naive} & 17.557 & 61.533 & 488.208& 4680.881 & 7.74 & 9.186 & 11.556& 23.832 & 0.914 & 9.135 & 91.346& 913.464\\
 & \multirow{-2}{*}{ \sclaz 40\%} & \textsc{occp} & 174.428 & 308.299 & 496.166& 2402.201 & 48.204 & 51.218 & 73.566& 80.282 & 0.411 & 4.518 & 74.752& 821.991\\
 &  & \textsc{naive} & 19.333 & 57.304 & 435.47& 4041.03 & 7.399 & 9.786 & 11.244& 23.329 & 1.523 & 15.225 & 152.244& 1522.44\\
\multirow{-12}{*}{ \bplaz} & \multirow{-2}{*}{ \scmal} & \textsc{occp} & 28.863 & 30.14 & 50.095& 249.054 & 15.073 & 15.047 & 15.061& 15.078 & 0.076 & 0.761 & 7.612& 76.122\\
\midrule
 &  & \textsc{naive} & 14.892 & 31.371 & 181.337& 9360.574 & 7.132 & 7.959 & 11.731& 47.969 & 1.736 & 17.331 & 173.285& 1732.823\\
 & \multirow{-2}{*}{ \schap} & \textsc{occp} & 28.851 & 29.997 & 52.597& 268.584 & 15.541 & 15.568 & 15.531& 15.623 & 0.087 & 0.867 & 8.664& 86.641\\
 &  & \textsc{naive} & 15.182 & 33.066 & 168.055& 8552.99 & 6.87 & 9.055 & 16.086& 176.006 & 1.562 & 17.331 & 173.285& 1732.823\\
 & \multirow{-2}{*}{ \sclaz 10\%} & \textsc{occp} & 138.944 & 121.621 & 108.981& 466.942 & 24.588 & 20.746 & 22.115& 21.373 & 0.128 & 1.324 & 14.209& 158.409\\
 &  & \textsc{naive} & 19.11 & 81.091 & 689.727& 6714.39 & 7.086 & 8.429 & 13.649& 54.558 & 1.388 & 13.865 & 138.628& 1386.258\\
 & \multirow{-2}{*}{ \sclaz 20\%} & \textsc{occp} & 76.496 & 99.202 & 137.277& 786.072 & 30.669 & 26.818 & 28.136& 31.414 & 0.192 & 1.922 & 23.206& 287.504\\
 &  & \textsc{naive} & 18.884 & 73.556 & 604.229& 5866.896 & 6.813 & 8.968 & 12.94& 52.778 & 1.215 & 12.132 & 121.299& 1212.976\\
 & \multirow{-2}{*}{ \sclaz 30\%} & \textsc{occp} & 96.825 & 137.232 & 186.412& 1500.339 & 36.262 & 32.892 & 34.629& 52.303 & 0.279 & 2.649 & 32.115& 540.063\\
 &  & \textsc{naive} & 17.574 & 65.538 & 520.883& 5022.172 & 6.789 & 9.382 & 13.629& 54.728 & 1.041 & 10.399 & 103.971& 1039.694\\
 & \multirow{-2}{*}{ \sclaz 40\%} & \textsc{occp} & 203.252 & 215.02 & 383.093& 3472.424 & 48.207 & 39.874 & 62.695& 103.02 & 0.463 & 3.502 & 67.985& 1224.239\\
 &  & \textsc{naive} & 19.444 & 61.271 & 469.616& 11951.525 & 5.076 & 7.344 & 11.566& 52.424 & 1.736 & 17.331 & 173.285& 1732.823\\
\multirow{-12}{*}{ \bpmat} & \multirow{-2}{*}{ \scmal} & \textsc{occp} & 28.779 & 30.245 & 52.138& 273.911 & 15.075 & 15.095 & 15.065& 15.153 & 0.087 & 0.867 & 8.664& 86.641\\
\midrule
 &  & \textsc{naive} & 15.109 & 33.37 & 210.528& 1870.644 & 6.646 & 7.548 & 12.781& 63.966 & 1.997 & 19.944 & 199.417& 1994.143\\
 & \multirow{-2}{*}{ \schap} & \textsc{occp} & 29.027 & 32.121 & 56.457& 308.65 & 15.591 & 15.571 & 15.583& 15.6 & 0.1 & 0.997 & 9.971& 99.707\\
 &  & \textsc{naive} & 15.188 & 33.062 & 195.587& 10848.192 & 6.362 & 8.434 & 17.893& 271.731 & 1.797 & 19.213 & 199.417& 1994.143\\
 & \multirow{-2}{*}{ \sclaz 10\%} & \textsc{occp} & 89.576 & 84.419 & 110.814& 543.501 & 25.549 & 22.548 & 22.441& 21.919 & 0.149 & 1.539 & 17.665& 188.28\\
 &  & \textsc{naive} & 20.431 & 91.117 & 797.028& 7731.012 & 6.582 & 7.999 & 14.752& 73.555 & 1.598 & 15.956 & 159.534& 1595.314\\
 & \multirow{-2}{*}{ \sclaz 20\%} & \textsc{occp} & 65.077 & 89.545 & 164.705& 762.991 & 31.662 & 27.73 & 28.746& 28.903 & 0.219 & 2.255 & 27.852& 286.492\\
 &  & \textsc{naive} & 19.59 & 82.883 & 700.215& 6756.859 & 6.284 & 8.507 & 14.094& 70.257 & 1.398 & 13.961 & 139.592& 1395.9\\
 & \multirow{-2}{*}{ \sclaz 30\%} & \textsc{occp} & 96.028 & 125.428 & 235.559& 1496.706 & 34.495 & 37.188 & 37.97& 46.337 & 0.276 & 3.743 & 43.423& 535.427\\
 &  & \textsc{naive} & 17.566 & 73.536 & 605.565& 5804.2 & 6.291 & 8.897 & 14.835& 73.48 & 1.198 & 11.967 & 119.65& 1196.486\\
 & \multirow{-2}{*}{ \sclaz 40\%} & \textsc{occp} & 167.115 & 210.084 & 334.589& 2857.447 & 44.756 & 45.045 & 51.549& 78.539 & 0.48 & 5.023 & 59.526& 1007.042\\
 &  & \textsc{naive} & 19.16 & 67.599 & 548.21& 5050.571 & 5.283 & 7.733 & 13.606& 71.242 & 1.997 & 19.944 & 199.417& 1994.143\\
\multirow{-12}{*}{ \bpmer} & \multirow{-2}{*}{ \scmal} & \textsc{occp} & 28.904 & 32.997 & 56.712& 312.064 & 15.123 & 15.1 & 15.119& 15.13 & 0.1 & 0.997 & 9.971& 99.707\\
\midrule
 &  & \textsc{naive} & 15.89 & 33.291 & 200.33& 1764.919 & 8.421 & 9.002 & 10.21& 21.002 & 1.992 & 19.923 & 199.232& 1992.32\\
 & \multirow{-2}{*}{ \schap} & \textsc{occp} & 30.448 & 32.437 & 54.457& 286.605 & 15.57 & 15.596 & 15.582& 15.596 & 0.1 & 0.996 & 9.962& 99.616\\
 &  & \textsc{naive} & 16.628 & 33.204 & 188.464& 10390.837 & 8.189 & 10.076 & 14.414& 88.713 & 1.793 & 19.326 & 199.232& 1992.32\\
 & \multirow{-2}{*}{ \sclaz 10\%} & \textsc{occp} & 92.283 & 117.923 & 121.541& 510.942 & 25.393 & 22.287 & 22.053& 22.195 & 0.143 & 1.546 & 17.117& 189.602\\
 &  & \textsc{naive} & 21.119 & 90.42 & 763.731& 7422.313 & 8.37 & 9.586 & 11.837& 24.066 & 1.594 & 15.939 & 159.386& 1593.856\\
 & \multirow{-2}{*}{ \sclaz 20\%} & \textsc{occp} & 55.789 & 92.639 & 141.294& 714.371 & 30.874 & 25.818 & 27.783& 28.804 & 0.198 & 2.065 & 24.821& 294.531\\
 &  & \textsc{naive} & 19.58 & 81.571 & 670.879& 6510.869 & 8.054 & 10.141 & 11.268& 23.06 & 1.395 & 13.946 & 139.462& 1394.624\\
 & \multirow{-2}{*}{ \sclaz 30\%} & \textsc{occp} & 92.422 & 118.021 & 215.833& 1719.338 & 34.328 & 36.16 & 36.906& 55.015 & 0.292 & 3.697 & 40.012& 664.937\\
 &  & \textsc{naive} & 19.555 & 71.534 & 578.225& 5576.213 & 8.07 & 10.689 & 11.882& 24.199 & 1.195 & 11.954 & 119.539& 1195.392\\
 & \multirow{-2}{*}{ \sclaz 40\%} & \textsc{occp} & 142.516 & 221.83 & 405.902& 4516.465 & 41.16 & 47.013 & 61.767& 119.669 & 0.408 & 5.281 & 76.688& 1707.751\\
 &  & \textsc{naive} & 19.108 & 65.873 & 518.917& 4762.235 & 6.987 & 9.396 & 10.761& 22.846 & 1.992 & 19.923 & 199.232& 1992.32\\
\multirow{-12}{*}{ \bpspf} & \multirow{-2}{*}{ \scmal} & \textsc{occp} & 30.438 & 32.477 & 54.372& 293.116 & 15.104 & 15.125 & 15.113& 15.127 & 0.1 & 0.996 & 9.962& 99.616\\

\bottomrule
\end{tabular}

  }
\end{table*}

\begin{custombox}[\req{3} -- In summary]
Our experiments demonstrate that adopting an informed step size improves time requirements and gas costs for longer-running benchmark problems, particularly when scaled by a factor of \num{1000}. 
Notably, while the performance gains become most evident after reaching a scale factor of \num{1000}, some improvements are already observed for time at smaller scales, such as \num{100}. 
However, at this lower threshold, reductions in gas costs remain negligible compared to the baseline.
Expectedly, the informed step size consistently reduces the number of re-executed expressions across all tested scales (similar to \req{2}).

Compared to the non-informed variation, time requirements are reduced by up to 26-fold for a step size of \num{100} and 10-fold for \num{1000}. 
Similarly, gas costs are reduced by up to 44-fold for a step size of \num{100} and 6-fold for \num{1000}. 
\end{custombox}

\section{Related Work}
\label{sec:related_work}

In this section, we provide a brief overview of prior works on verification and computational integrity approaches that relate to our proposed approach.

\paragraph{Hardware-based verification}
Hardware-based verification can offer certain guarantees regarding the authenticity and integrity of an application.
However, this approach relies on specialized hardware, such as \ac{sgx}~\cite{costan_intel_2016}, \ac{kaplan_amd_2016}~\cite{kaplan_amd_2016}, or \ac{trustZone}~\cite{pinto_demystifying_2019}, which can be costly and not always available.
Additionally, there is still the possibility that malicious actors may alter the results after execution, as highlighted in the challenges of \ac{sgx}~\cite{fei_security_2021}.

\paragraph{Constraint solvers}
They translate code into constraints, which can only be solved if all the arguments provide a solution to the equation system of constraints.
Systems such as \textsc{Piperine}~\cite{lee_replicated_2020}, \textsc{Pantry}~\cite{braun_verifying_2013}, and \textsc{Spice}~\cite{setty_proving_2018} provide such functionality.
However, all the aforementioned systems are limited by one or more of the following characteristics:
\begin{inparaenum}
    \item they require changes to the code to work,
    \item have a fixed bound for loops,
    \item static size for data structures (\eg, fixed tree depth), and
    \item high latency due to waits between batches of verifications.
\end{inparaenum}
While these limitations may not be problematic for smaller applications or toy examples, they can hinder the performance and scalability of larger applications with considerable dataset sizes.
Additionally, the verification process would require either a deep understanding of the code's internals or blind trust in a third party that provides the constraints.

\paragraph{Software-based verification}
Sasson \etal introduced \textsc{zk-SNARKs}, a software-based verification approach for non-interactive zero-knowledge proofs.
The approach is based on arithmetic circuits and assertions that do not require a re-execution to verify~\cite{ben-sasson_succinct_2014}.
Later, Sasson \etal improved their previous approach and introduced \textsc{zk-STARKs} that introduce transparency for zero-knowledge systems.
The approach uses arithmetic circuits to generate proofs~\cite{ben-sasson_scalable_2018}.
Bitansky \etal published \textsc{SNARGs} (Succinct Non-Interactive Arguments) that improve the required complexity for verifying based on cryptographic transformations, therefore, reducing the verification time for software-based verification systems~\cite{bitansky_succinct_2022}.

However, all of these approaches rely on the ability to compile code to circuits to verify that a computation was correct, where a \enquote{prover} has to execute and generate a transcript (proof), which is then verified by the \enquote{verifier}~\cite{walfish_verifying_2015}.
However, most such approaches require that computations are bounded.
In other words, circuits have a fixed size.
Consequently, most software-based verification systems~\cite{zeiselmair_analysis_2021} can not handle input-dependent loops, recursions, and dynamic-sized data structures (\eg, depth for trees).
Therefore, such systems are not well suited for larger (real-world) applications, as seen in most industrial applications.
To reiterate, a deep understanding of the internals of the code is a prerequisite for approaches based on writing assertions to verify that an execution was executed correctly.
Additionally, writing non-trivial assertions requires time and effort.
Therefore, such systems can not easily be applied to our use cases.

\paragraph{On-Chain verification}
Teutsch and Reitwießner proposed \textsc{TrueBit}, a scalable verification solution for blockchains that is based on an elaborate incentive system.
In this approach, a solver executes the full program, and in case of a dispute, is challenged and must play a verification game to prove that the solution was correct~\cite{teutsch_scalable_2019}.
However, as discussed earlier, re-executing the full program significantly reduces the usability of such a system due to the increase in computation and time requirements.

\section{Conclusions and Future Work}
\label{sec:conclusions}

We have demonstrated the feasibility of our proposed \acf{mi} and \acf{occp} for certifying program executions with the help of a layer 2 blockchain.
Our experimental results indicate that our \ac{mi} prototype can accurately reproduce correct results for every step size.
However, a trade-off between trust and performance exists, which requires further investigation to determine an ideal balance between the two. 
On average the step size increase from \num{100} to \num{1000} speeds up the certification process by a factor of \num{7.371} while only slightly increasing the number of executed expressions in certain scenarios.
In-depth experiments and optimizations are needed to mitigate the impact of a lower step size.

Additionally, our proposed \ac{occp} was able to certify correct executions, while outperforming the baseline, which re-executes the full program multiple times, with fewer executed expressions in all proposed scenarios.
Furthermore, our proposed approach allows for the detection of incorrect or malicious actions with similar effort as certifying a correct one.

In the future, we plan to investigate the adaptation of the mechanisms devised for \ac{mi} to other programming languages, providing them with the feature of segmentation and replay.
While this paper presents evidence on the feasibility of the approach, a large-scale case study is required to thoroughly examine how well this approach scales to real-world scenarios, 
considering both the performance impact and the practical challenges of deploying it across diverse, complex environments.

For \ac{occp}, we aim to explore the use of different blockchains and develop an incentive mechanism alongside a reward system that better defines the roles and compensations for participants solving the proposed problem. 
Specifically, potential incentive mechanisms could include a paid service model, funded either by universities and journals or by users who wish to verify their execution, paying into a smart contract. 
This smart contract would then fairly distribute monetary remuneration to workers based on their contributions. 
Additionally, to further ensure reliable and non-malicious participation, we propose a dual-layer approach: 
\begin{inparaenum}
\item integrating a mechanism to detect malicious workers, thereby encouraging trustworthy behavior, and 
\item requiring workers to deposit a small stake into the smart contract.
\end{inparaenum} 
Workers would forfeit this deposit in cases of malicious behavior, but non-malicious participants would receive it back, providing both a deterrent against misconduct and an additional incentive to act responsibly.

Future work can explore extending our approach to multi-threading aspects. Adding a sequentially numbered annotation layer, using the seqId strategy, to track the evaluation progress of threads within its assigned code derivation. This would capture a memory snapshot showing all active threads at a given time. 
Moreover, future work could focus on enhancing the performance of the \ac{mi} by exploring alternative Intermediate Representations (IRs) beyond the current \ac{ast} approach, aiming to achieve faster verification times and support more advanced optimizations.
The proposed certification protocol can succeed if the likelihood of reaching a snapshot with output altered by a concurrent side effect remains statistically probable within expected retries. The annotation layer would help manage multiple threads, allowing computation to continue from where each thread left off.
If concurrency doesn't affect the snapshot's memory state, the snapshot is verified. Otherwise, the same concurrency conditions must be reproducible through a quorum. 
Further analysis is needed to define an acceptable quorum and the probability of reaching the same memory state. Future research can also focus on a concurrency-aware snapshotting strategy to limit side effects and maintain consistent outputs.

In order to address non-determinism, we plan to explore the feasibility of using a sampling mechanism for random generators until a confidence threshold is reached. 
This would introduce an element of uncertainty, making it crucial to carefully analyze the impact on system behavior and overall reliability.

Additionally, future work can focus on implementing an identity mechanism to prevent address spoofing and flooding by malicious workers
 by employing \ac{did} frameworks, each worker would possess an identity anchored in a public blockchain, where cryptographic signatures ensure authenticity without relying on a centralized authority. 
This would enable consistent worker recognition, prevent malicious actors from forging multiple identities. 
Additionally, verifiable credentials could be used to certify a worker’s qualifications and past performance while preserving privacy, further mitigating risks of impersonation or malicious behavior. 
Future iterations of our system could implement these mechanisms as a service to strengthen security.

Lastly, a non-local adaptation of \ac{ipfs} should be analyzed and evaluated in future iterations as well as the use of consistent hashing to improve performance of our approach.

In conclusion, our proposed \ac{mi} and \ac{occp} hold promise in enhancing the trustworthiness and security of program executions through segmentation and certification, respectively.
It encourages further investigation and development in this area by the research community.

\section{Limitations and Threats to Validity}
\label{sec:experiments:threats}
\paragraph{Malicious worker scenarios} 
Our experiments specifically evaluate the scenario where malicious workers intentionally provides incorrect results. 
In practice, malicious actors may attempt to interfere with the certification process in other ways, such as by withholding results or intentionally producing conflicting outputs.
Consequently, the robustness of our protocol against other types of malicious interference remains uncertain.

\paragraph{Collusion between malicious users and workers}
We did not explore scenarios involving collusion between malicious workers and users. 
If more than 50\% of the workers are malicious, they could potentially disrupt the certification process, leading to failure of the protocol and the underlying blockchain. 
This requires deeper analysis, which is out of scope for the current work.

\paragraph{Local blockchain evaluation}
The use of a local blockchain platform for our experiments may not fully represent the performance of our protocol in non-local or private blockchain environments. 
Additionally, the use of a local \emph{Amazon S3} instance for trace data storage might impact protocol performance, potentially affecting the performance of the protocol.

\paragraph{Majority voting mechanisms}
Our protocol relies on majority voting for certification.
While it is designed to detect discrepancies between executions, it assumes that the majority of participants are honest. 
Collusion among malicious workers and malicious users could undermine the voting mechanism's effectiveness.
However, this is an intrinsic limitation of majority voting mechanisms.

\paragraph{Generalizability to other languages and paradigms}
Additionally, we evaluated the feasibility of our proposed programming language on specific use cases (see \cref{subsec:rq1Exp}).
Further work is required to extend our work to other programming languages and evaluate real-world applications.

\paragraph{Task distribution fairness and integrity}
Our current implementation does not address fairness or integrity in task distribution to workers. 
We mitigate this partially by tracking tasks and worker identification to prevent disputes from being assigned to the same worker. 
However, more robust mechanisms need to be developed to enhance fairness.

\paragraph{Lack of identity mechanisms}
Although our implementation includes unique identifiers for workers, it currently lacks a robust identity verification mechanism.
To address this limitation, future work could integrate \ac{did} frameworks, as discussed in prior studies~\cite{soltani2018HyperLedger,bouras2020DLTeHealth,liu2020BCidentitySystems,soltani2021SurveySSI}. 
Specifically, platforms like Hyperledger Indy~\cite{hyperledger_indy} or uPort~\cite{lundkvist_uport_nodate} could provide a foundation for decentralized and cryptographically verifiable identities.
By using \ac{did}, workers would be associated with cryptographically secure and verifiable identities, ensuring consistent worker recognition and preventing malicious actors from forging multiple identities.

\paragraph{Smart contract optimization}
We also acknowledge that our implementation of the smart contract may not be optimal, thus leading to higher gas consumption.
However, this requires further research into the optimization of smart contracts and is out of scope for this paper.

\paragraph{False positives vs. false negatives}
Finally, it is worth noting that our protocol is designed to be more resilient to false positives (\ie, falsely certifying a task as correct) than false negatives (\ie, failing to certify a correct task).
As a result, the protocol may require tasks to be re-executed if they fail to produce a certificate, even if they are correct.
This may lead to additional computational overhead and delay in some scenarios.

\section*{Acknowledgments}

The research leading to these results has been supported by the Swiss National Science Foundation (SNSF) through Grant SNSF204632.

We sincerely thank Dr. Pooja Rani for her invaluable guidance and efforts in helping us revise and improve our paper.

\balance
\bibliography{references, urls}

\clearpage
\begin{complete-version}
\nobalance
\begin{IEEEbiography}%
    [{\includegraphics[width=1in,height=1.25in,clip,keepaspectratio]{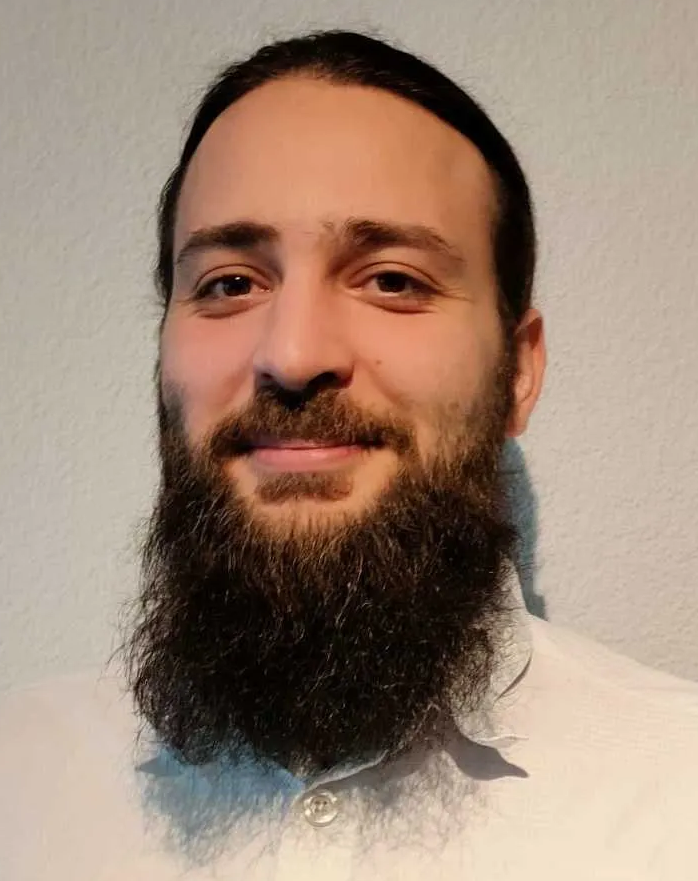}}]
    {Alex Wolf}%
    is a doctoral candidate in the Software Evolution and Architecture Lab (s.e.a.l.) at the University of Zurich, Switzerland,
    where their research focuses on advancing machine learning, software architecture, and engineering.
    He earned a Master's degree in Software Systems from the University of Zürich in 2022.
    Prior to beginning their doctoral studies, Alex gained industry experience through various software engineering roles,
    providing a strong foundation in practical and technical problem-solving. Their research interests lie at the intersection of
    machine learning and software engineering, with a particular focus on practical applications and the integration of blockchain technology.
    Contact him at \href{mailto:wolf@ifi.uzh.ch}{wolf@ifi.uzh.ch}.
\end{IEEEbiography}

\begin{IEEEbiography}%
    [{\includegraphics[width=1in,height=1.25in,clip,keepaspectratio]{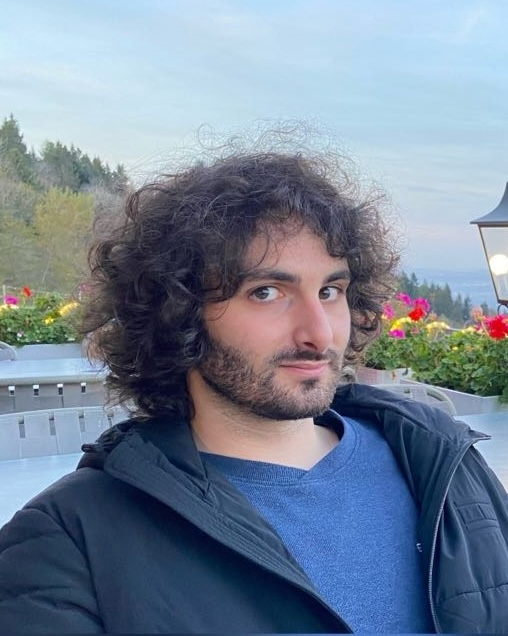}}]
    {Marco Edoardo Palma}%
    is a doctoral candidate in the Software Evolution and Architecture Lab (s.e.a.l.) at the University of Zurich, Switzerland.
    He earned a First-Class Honours degree in Computer Science with Artificial Intelligence from the University of Southampton, UK, in 2021.
    His research explores the development of artificial intelligence tools and strategies to enhance software engineering processes and tools.
    Currently, his work centres on the Deep Abstraction strategy, which automatically compiles algorithms with high space and time complexity
    into efficient statistical models.
    Contact him at \href{mailto:marcoepalma@ifi.uzh.ch}{marcoepalma@ifi.uzh.ch}.
\end{IEEEbiography}

\begin{IEEEbiography}%
    [{\includegraphics[width=1in,height=1.25in,clip,keepaspectratio]{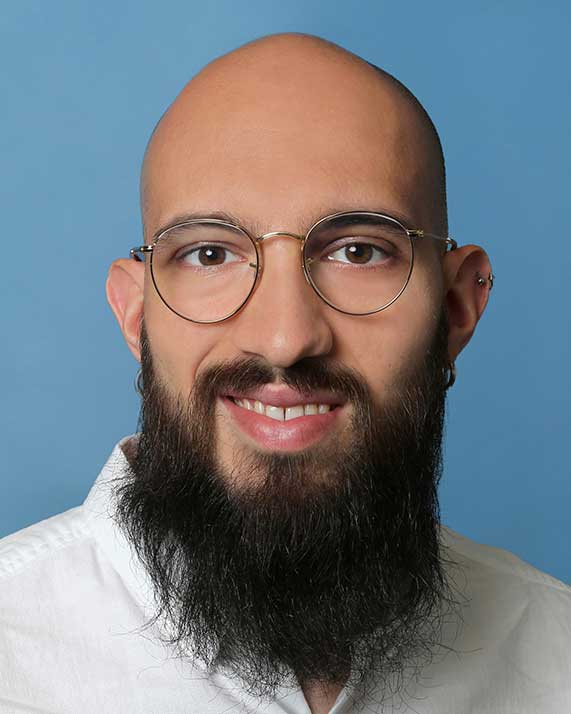}}]
    {Pasquale Salza}%
    is a Senior Research Associate
    in the Software Evolution and Architecture Lab (s.e.a.l.)
    at the University of Zurich, Switzerland.
    He received a Ph.D. degree in Computer Science from the University of Salerno, Italy.
    His research interests include software engineering, machine learning, cloud computing, and evolutionary computation.
    Contact him at \href{mailto:salza@ifi.uzh.ch}{salza@ifi.uzh.ch}.
\end{IEEEbiography}

\begin{IEEEbiography}%
    [{\includegraphics[width=1in,height=1.25in,clip,keepaspectratio]{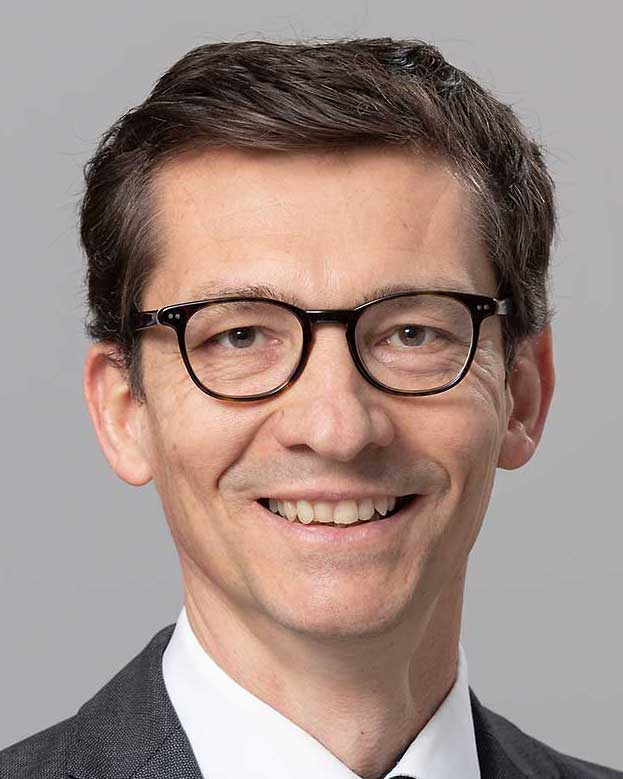}}]
    {Harald C. Gall}%
    is professor of software engineering and director of the Software Evolution and Architecture Lab (s.e.a.l.)
    in the Department of Informatics at the University of Zurich, Switzerland.
    He held visiting positions at Microsoft Research, USA, and University of Washington in Seattle, USA.
    His research interests are software evolution, software architecture, software quality, and green
    software engineering. He has worked on developing new ways in which data mining of software repositories
    and machine learning can contribute to a better understanding and improvement of software development.
    Contact him at \href{mailto:gall@ifi.uzh.ch}{gall@ifi.uzh.ch}.
\end{IEEEbiography}

\vfill

\end{complete-version}

\end{document}